\documentclass[sigconf]{acmart}

\AtBeginDocument{%
  }

\setcopyright{acmlicensed}
\copyrightyear{2018}
\acmYear{2018}
\acmDOI{XXXXXXX.XXXXXXX}

\acmConference[Conference acronym 'XX]{Make sure to enter the correct
  conference title from your rights confirmation emai}{June 03--05,
  2018}{Woodstock, NY}
\acmISBN{978-1-4503-XXXX-X/18/06}

\usepackage{tikz}
\usepackage{amsmath}
\usepackage{placeins}
\usepackage{color}
\usepackage{listings}
\usepackage{xcolor}
\usepackage{filecontents}
\usepackage{graphicx}
\usepackage{subcaption}
\usepackage{amsmath}
\usepackage{listings}
\usepackage{booktabs} 
\usepackage{multicol} 
\usepackage{multirow} 
\usepackage{amsmath}  
\usepackage[linesnumbered,ruled,vlined]{algorithm2e}
\usepackage{graphicx}
\usepackage{multirow}
\usepackage{xcolor}
\usepackage{diagbox} 
\usepackage[]{hyperref}
\usepackage{graphicx}
\usepackage{caption}
\usepackage{subcaption}
\usepackage{makecell}  

\usepackage{enumitem}  
\usepackage{array}    

\usepackage{balance}
\usepackage[utf8]{inputenc}
\usepackage{algorithmicx}
\usepackage{algpseudocode}  
\usepackage{setspace}

\newcommand{\xq}[1]{{\color{black} #1}}

\newcommand{\lei}[1]{\textcolor{red}{Lei: #1}}
\renewcommand{\lei}[1]{}
\begin{document}

\setcopyright{none}
\renewcommand\footnotetextcopyrightpermission[1]{} 
\pagestyle{plain} 
\settopmatter{printfolios=true,printccs=false,printacmref=false}

\date{}

\title{HARMONY: A Scalable Distributed Vector Database for High-Throughput Approximate Nearest Neighbor Search}

\newcommand{\cred}[1]{\textcolor{red}{#1}}
\newcommand{\zf}[1]{\textcolor{violet}{#1}}

\author{
Qian Xu\textsuperscript{1},
Feng Zhang\textsuperscript{1},
Chengxi Li\textsuperscript{1},
Zheng Chen\textsuperscript{1},
Lei Cao\textsuperscript{2},
Jidong Zhai\textsuperscript{3},
Xiaoyong Du\textsuperscript{1}, \\
\textsuperscript{1}Renmin University of China \quad
\textsuperscript{2}Massachusetts Institute of Technology \quad
\textsuperscript{3}Tsinghua University
}

\begin{abstract}
Approximate Nearest Neighbor Search (ANNS) is essential for various data-intensive applications, including recommendation systems, image retrieval, and machine learning.
Scaling ANNS to handle billions of high-dimensional vectors on a single machine presents significant challenges in memory capacity and processing efficiency. 
To address these challenges, distributed vector databases leverage multiple nodes for the parallel storage and processing of vectors. 
However, existing solutions often suffer from load imbalance and high communication overhead, primarily due to traditional partition strategies that fail to effectively distribute the workload. 
In this paper, we introduce Harmony, a distributed ANNS system that employs a novel multi-granularity partition strategy, combining dimension-based and vector-based partition.
This strategy ensures a balanced distribution of computational load across all nodes while effectively minimizing communication costs. 
Furthermore, Harmony incorporates an early-stop pruning mechanism that leverages the monotonicity of distance computations in dimension-based partition, resulting in significant reductions in both computational and communication overhead. 
We conducted extensive experiments on diverse real-world datasets, demonstrating that Harmony outperforms leading distributed vector databases, achieving 4.63$\times$ throughput on average in four nodes and 58\% performance improvement over traditional distribution for skewed workloads.
\end{abstract}



\maketitle 
\pagestyle{plain} 

\def\sysname{Harmony}
\newcommand{\deletexq}[1]{{\color{orange}deletexq: #1}}



\section{Introduction}
\label{sec:intro}

Approximate nearest neighbor search (ANNS) is at the core of many data-intensive applications such as search engines~\cite{li2021a_SOSP_35,Spann,kulis2009a_SOSP_32}, e-commerce platforms~\cite{li2018a_SOSP_34}, recommendation systems~\cite{koenigstein2012a_OSDI_49,li2017a_OSDI_53,lian2020a_OSDI_56,xiao2022a_OSDI_84} and natural language processing tasks~\cite{mikolov2013a_SOSP_42,pennington2014a_SOSP_49}\lei{such as RAG ...}. 
Despite significant advancements in indexing structures and query algorithms~\cite{ADSampling,HNSW}, over 90\% of the time spent on ANNS is dedicated to computing distances between vectors in cluster-based indices~\cite{ADSampling}. \lei{So we specifically target cluster-based indices, such as product quantization? However, distance computation in PQ seems not super expensive?}
Moreover, the volume of unstructured data, which often must be converted to high dimensional vectors before analysis, continues to increase. For instance, during Alibaba’s shopping festival, 500PB of data is generated daily~\cite{gabriel1969a_SOSP_20}, while YouTube receives over 500 hours of content every minute~\cite{youtube}. \lei{Why is this relevant to vector DB? Do they have to convert this amount of data to vectors?} 
This escalating demand highlights the need for distributed solutions to store vectors and process ANNS queries in parallel across multiple nodes.
Distributed computing has achieved remarkable success across various domains, including web-scale indexing~\cite{dean2008mapreduce} and big data analytics~\cite{white2012hadoop,zaharia2010spark}. 
Building on these successes, distributed vector databases~\cite{NSDI2023,wang2021a_SOSP_65} have emerged as a promising approach to accelerate large-scale ANNS. However, the high dimensionality of vectors and the need for full-dimensional distance computations make conventional partitioning methods unsuitable. \xq{Moreover, the popular graph-based segmentation in standalone machines is not well compatible with distributed features, as query paths for vectors tend to introduce edges across machines, resulting in high latency.} \lei{such as? There are a bunch of partitioning methods. Some cannot handle high D data, but some may be OK} Thus, our work focuses on developing a distributed-friendly vector distribution method and optimizing distributed vector distance computations \xq{based on cluster-based vectors}. \lei{We need to more carefully analyze these existing methods, as partitioning seems the key of this work.}

\xq{Distributed vector databases provide benefits in both space efficiency and time overhead reduction.
Existing studies show that most time in Approximate Nearest Neighbor Search (ANNS) is spent on distance calculations between vectors. 
In cluster-based vectors\lei{again we don't really know what index is used}, distance computations accounted for 99.7\% of the total time when conducting searches with precision 99\% on the \emph{msong} \lei{need a reference} dataset~\cite{ADSampling}. 
Optimizing this time overhead can significantly enhance ANNS speed. 
Distributed vector databases can substantially reduce query latency by processing large ANNS workloads concurrently~\cite{wang2021a_SOSP_65}. 
On the other hand, distributing vectors across multiple machines alleviates memory pressure on individual nodes.
}
These advantages make distributed architectures an ideal solution for handling the increasing volume of vector data in modern applications.\lei{I am not sure if it is necessary to use so much space to sell the good of distributed vector DB, as we are not the first one to propose building a distributed vector DB. We want to emphasize the key challenge of partitioning if it is the key of the paper.}

Despite advancements, distributed acceleration for Approximate Nearest Neighbor Search (ANNS) still faces three key challenges. 
First, traditional partition methods, such as cluster-based or hash-based partition, often result in load imbalances in ANNS systems. These imbalances occur when certain machines handle disproportionately large computations, which degrade performance. An adaptive indexing scheme is required to resolve this issue and enhance ANNS efficiency.
Second, distributing queries across machines while minimizing communication and computation overhead remains challenging in distributed ANNS systems. Distributed vector databases often transmit intermediate results, which typically account for less than 10\% of the size of the original vectors~\cite{NSDI2023,distribute1}. Although smaller, the communication overhead remains significant due to the bandwidth disparity between transmission (up to 100Gb/s) and computation (hundreds of GB/s), causing delays\cite{distribute1}. Simple query distribution strategies can cause bottlenecks and degrade performance, especially under imbalanced workloads.
Third, inter-machine pruning offers opportunities to reduce redundant computations in ANNS, but designing lightweight pipelines for pruning with minimal overhead is a major challenge in improving query efficiency.

Previous research on distributed vector databases has focused on two main aspects: (1) improving query performance in multi-machine settings~\cite{distribute1, distribute2, distribute3}, and (2) ensuring the accuracy of query results across multiple nodes~\cite{distribute4, guo2022manu}.  
However, these studies do not address performance guarantees under load imbalances or explore pruning strategies that reduce inter-machine workload. In addition to coordinating machines, inter-machine pruning can also reduce workload and balance imbalanced loads, ultimately improving throughput.


\begin{figure}
    \centering\includegraphics[width=\linewidth,height=0.7\linewidth]{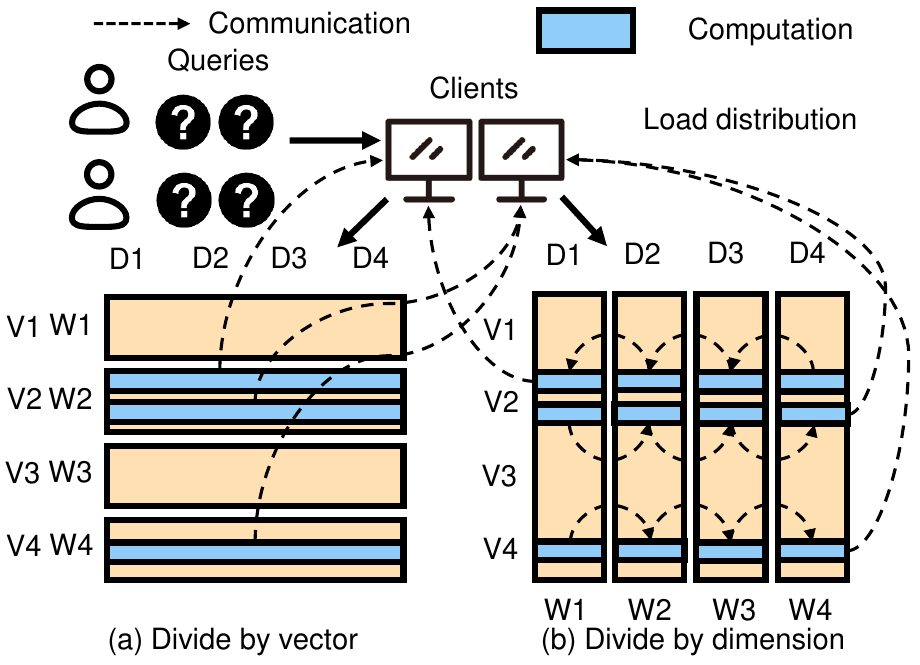}

    \caption{An example of overhead in different partition granularity when searching. Where V\_i denotes the ith block according to vector-based partition and D\_j denotes the jth block according to dimension-based partition.}
    \label{fig:intro}
\end{figure}

\xq{We have made two key observations. First, vectors can be partitioned at various granularities, such as vector-based and dimension-based partition, as shown in Figure~\ref{fig:intro}. Each strategy has distinct computational and transmission loads, with W1-W4 representing different computing machines. Specifically, traditional vector-based partition~\cite{NSDI2023,wang2021a_SOSP_65}} involves dividing the dataset by assigning entire vectors to different nodes (V1-V4 in Figure~\ref{fig:intro}~(a)). This approach simplifies distance computation, with each node communicating with the master node only before and after the calculation. However, it often causes load imbalance, with some nodes becoming overloaded with ``hot'' partitions, while others remain underutilized, particularly under uneven query workloads. For example, as shown in Figure~\ref{fig:intro}~(a), in the case of an unbalanced load, only W2 and W4 perform the distance computation. 
In contrast, as shown in Figure~\ref{fig:intro}~(b), dimension-based partition distributes the dataset by splitting the dimensional space across nodes, with each node handling a subset of the vector dimensions (D1-D4 in Figure~\ref{fig:intro}~(b)). This method ensures a balanced load across all nodes regardless of the query patterns, as each node contributes equally to the distance computations. However, it necessitates multiple communications between the client and the nodes for each query, increasing the communication overhead and potentially impacting query latency. \lei{Is this observation or the limitation of SOTA? What partitioning does the SOTA use? Vector based or dimension based? Or is dimension-based a new idea?}

Second, we find that dimension-based partition allows early stopping during distance computations. For instance, in Figure~\ref{fig:intro}(b), if the partial distance computed from W1 and W2 exceeds the current worst distance, the remaining computations in dimensions W3 and W4 can be safely skipped. This early termination enables pipelined processing and pruning techniques, significantly reducing computational burden and accelerating query responses.

Based on these observations, we present \sysname{}, a \underline{\textbf{h}}igh throughput \underline{\textbf{a}}nd \underline{\textbf{r}}obust \underline{\textbf{m}}ulti-node \underline{\textbf{o}}ptimized \underline{\textbf{n}}earest neighbor search s\underline{\textbf{y}}stem. It introduces three main innovations. 
\xq{First, \sysname{} integrates vector-based and dimension-based partition into a hybrid index construction framework. This approach, guided by a cost model, dynamically adapts to the characteristics of each query and the current cluster state, ensuring efficient query processing and stable performance under various loads.} \lei{The above sentence is hard to understand. a mixed solution that combines vector-based and dimension-based partitioning?} 
Second, \sysname{} addresses load imbalance by combining dimension-based and vector-based partition in a hybrid granularity scheme, ensuring efficient load distribution and maintaining high throughput for balanced workloads.
Third, Harmony introduces a pruning technique based on the monotonicity of distance computations in dimension-based partition. When partial vector dimensions no longer contribute to the final result, redundant calculations are skipped. This early pruning, coupled with a pipelined execution model, minimizes unnecessary computations and communications, especially for high-dimensional queries. \lei{pruning the query process? Shortcut the query process or early stopping?}

Our experimental evaluation on multiple real-world datasets shows that Harmony achieves substantial performance gains over state-of-the-art vector databases, with a 4.63$\times$ throughput improvement on average in four nodes and a 58\% improvement over traditional distribution for skewed loads.
We summarize our main contributions as follows:

\begin{itemize}[noitemsep,nolistsep,leftmargin=0.2in]
    \item We observe that dimension-based partition enables an even distribution of computational demands. Integrating it with vector-based partition, we propose a multi-granularity approach that optimizes resource allocation across nodes, improving system efficiency.
    \item We find that dimension-based partition allows early stopping during distance computations. We design a pruning strategy that reduces computational and communication overhead by utilizing intermediate results at the dimension level across nodes.
    \item We conduct comprehensive experiments demonstrating that, compared to the open-source system Faiss, Harmony achieves a 4.63$\times$ throughput improvement on average in four-node.
\end{itemize}

 \vspace{-0.1in}

\section{Background and Related Work}
\label{sec:back}

\subsection{Vector Database Systems}

With the rise of deep learning, vectors are now integral to a wide range of applications, including search engines~\cite{li2021a_SOSP_35,Spann,kulis2009a_SOSP_32}, e-commerce platforms~\cite{li2018a_SOSP_34}, recommendation systems~\cite{koenigstein2012a_OSDI_49,li2017a_OSDI_53,lian2020a_OSDI_56,xiao2022a_OSDI_84}, and natural language processing tasks~\cite{mikolov2013a_SOSP_42,pennington2014a_SOSP_49}. Vectors serve as representations for diverse data types, such as videos, texts, images, and audio. To enhance the performance of these applications, efficient vector search becomes crucial, driving the development of optimized vector databases~\cite{DBLP:conf/osdi/ZhangXCSXCCH00Y23_OSDI_0,DBLP:conf/sosp/XuLLXCZLYYYCY23_SOSP_0,DBLP:conf/ppopp23,HNSW,NSG,FPGA,RaBitQ}.

Vector databases face considerable challenges due to the rapid expansion of vector data. For example, the UCI Machine Learning Repository~\cite{UCI} lists over 100 open-source datasets with dimensions exceeding 100. At the same time, the "online" nature of these systems~\cite{dean2013a_OSDI_31,gray2000a_OSDI_36,jalaparti2013a_OSDI_42,nishtala2013a_OSDI_64} demands that vector searches be completed within milliseconds. This strict latency requirement contrasts with the exact search algorithms used in relational databases~\cite{clarkson-a_OSDI_28}, where most vectors must be scanned to retrieve the necessary results.

To meet these performance demands, vector database systems employ indexes to efficiently locate vectors. These indexes generally fall into two categories: graph-based and partition-based. Graph-based indexes~\cite{delaunay1934a_SOSP_15,dong2011a_SOSP_16,NSG,gabriel1969a_SOSP_20,hajebi2011a_SOSP_23,HNSW,toussaint1980a_SOSP_60,wang2012a_SOSP_63,DBLP:conf/osdi/ZhangXCSXCCH00Y23_OSDI_0} construct a graph based on vector distances, where the query process involves a breadth-first search. Partition-based indexes, such as cluster-based~\cite{facebook2020a_SOSP_17,babenko2014a_OSDI_17,baranchuk2018a_OSDI_19,Spann,jegou2010a_OSDI_44,j2011a_OSDI_45,kalantidis2014a_OSDI_48,zhang2014a_OSDI_90,DBLP:conf/sosp/XuLLXCZLYYYCY23_SOSP_0}, hash-based~\cite{datar2004a_OSDI_30,jain2008a_OSDI_41,wang2018a_OSDI_79,ren2020a_SOSP_51,weiss2009a_OSDI_81}, and high-dimensional tree-based~\cite{bentley1975a_OSDI_23,liu2004a_OSDI_57,muja2014a_OSDI_62,wang2014a_OSDI_78} methods, divide vectors into partitions based on distance metrics, enabling faster identification of nearest clusters before searching within them.

Although these indexing techniques improve query speed, they incur additional storage overhead. Since full-dimensionality is necessary to compute vector distances accurately, reducing storage costs without resorting to lossy compression techniques such as quantization~\cite{facebook2020a_SOSP_17} remains a challenge. As a result, attention is shifting towards distributed vector Approximate Nearest Neighbor Search (ANNS) schemes to address these challenges.

\subsection{Distributed Solutions in Data Management} \label{subsec:existing_distributed_solutions}

Beyond specialized vector databases, numerous distributed data management systems in both graph and relational domains offer valuable insights into handling massive datasets. For instance, graph databases like Neo4j~\cite{Neo4j}, TigerGraph~\cite{TigerGraph}, and JanusGraph~\cite{JanusGraph} employ partitioning (or sharding) strategies that split large graphs across multiple machines, aiming to reduce cross-partition edges and data transfers during queries. Similarly, distributed relational databases such as Google Spanner~\cite{Spanner}, CockroachDB~\cite{CockroachDB}, and F1~\cite{F1} achieve horizontal scalability by sharding tables across commodity servers, enabling transactional consistency and parallel query execution. Distributed query engines like Presto~\cite{Presto}, Spark SQL~\cite{PrestoSparkSQL}, and Apache Drill~\cite{ApacheDrill} apply massive parallel processing (MPP) and in-memory computation to further reduce query latencies.

Although these systems primarily target structured or graph-shaped data, they share common design principles that are highly relevant to vector databases. First, they demonstrate the importance of robust data partitioning, where carefully chosen partitioning schemes minimize inter-node communication and balance workloads. Second, they underline the critical role of parallel processing frameworks in reducing query latency and scaling to large datasets. Third, they highlight potential challenges such as load imbalance and excessive data transfers, which can degrade performance if not properly managed. As high-dimensional vector data continues to grow in volume and complexity, leveraging these established techniques—while also addressing the unique aspects of vector data—presents an opportunity to build more efficient and scalable distributed vector database solutions~\cite{distribute1,distribute2,distribute3,distribute4,NSDI2023,Distribute_VB}.

\section{Motivation}
\label{sec:obser}

\subsection{Revisiting Search and Cost Distribution in Distributed Vector Databases}
\label{subsec:motivation}

\textbf{Motivation 1: Dimension-Level Pruning via Distance Monotonicity.}
\label{Motivation2:pruning_dim}
When vectors are partitioned by dimension across multiple nodes, each node $k$ stores a (disjoint) subset of dimensions $I_k \subseteq \{1,\dots,d\}$. Formally, a $d$-dimensional vector $\mathbf{q} = (q_1,q_2,\dots,q_d)$ is split into $M$ parts:
\[
  \mathbf{q}^{(k)} \;=\; (q_i)_{i \in I_k}, 
  \qquad
  \bigcup_{k=1}^{M} I_k = \{1,2,\dots,d\}, 
  \quad
  I_k \cap I_{k'} = \varnothing \ (k\neq k').
\]
This partitioning induces a \emph{partial result} on each node:
\[
  \text{Partial}_k(\mathbf{p}, \mathbf{q}) 
  \;=\; 
  f(\mathbf{p}^{(k)},\;\mathbf{q}^{(k)}),
\]
where $f$ is the local distance or similarity function restricted to the dimensions $I_k$. Below, we consider two common metrics: squared Euclidean distance and cosine similarity.

\paragraph{1) Euclidean Distance.}
For vectors $\mathbf{p}, \mathbf{q}\in\mathbb{R}^d$, the squared Euclidean distance is
\[
  D_{\text{Eucl}}^2(\mathbf{p},\mathbf{q}) 
  \;=\; 
  \sum_{k=1}^M
  \sum_{\,i\in I_k} (p_i - q_i)^2
  \;=\;
  \sum_{k=1}^M 
  D_k^2(\mathbf{p},\mathbf{q}),
\]
where 
\(
D_k^2(\mathbf{p},\mathbf{q}) = \sum_{\,i\in I_k} (p_i - q_i)^2.
\)
Hence, the \emph{total} distance is the sum of partial distances across $M$ nodes.

\paragraph{2) Cosine Similarity.}
For \emph{angle-based} retrieval (e.g., maximizing cosine similarity), one often works with the dot product
\[
  \mathbf{p}\cdot\mathbf{q} 
  \;=\;
  \sum_{k=1}^M 
  \sum_{\,i\in I_k} p_i\,q_i
  \;=\;
  \sum_{k=1}^M
  \alpha_k(\mathbf{p},\mathbf{q}),
\]
where 
\(
\alpha_k(\mathbf{p},\mathbf{q}) = \sum_{\,i\in I_k} p_i\,q_i.
\)
Cosine similarity can be seen as $\frac{\mathbf{p}\cdot \mathbf{q}}{\|\mathbf{p}\|\|\mathbf{q}\|}$, which reduces to a scaled version of the dot product if the vectors are pre-normalized.

Therefore, we can distribute vectors to different machines by dividing the dimensions for a richer distribution strategy.

\textbf{Dimension-level pruning.} A key advantage of dimension-based partitioning is the ability to exploit monotonicity in distance computations for early stopping. The partial results (e.g., partial sums or dot products) from previous nodes are aggregated to make centralized pruning decisions. As soon as the aggregate result indicates that the final distance cannot improve upon the current best candidate, the remaining nodes can safely skip their distance calculations.

For example, let $\mathbf{p}$ and $\mathbf{q}$ be $d$-dimensional vectors and suppose the dimensions are split into $M$ disjoint subsets $I_1, I_2, \dots, I_M$, each subset residing on one machine. The partial distance computed by machine $k$ and the cumulative partial distance after $k$ machines have finished is:
\[
d_k^2(\mathbf{p}, \mathbf{q}) 
\;=\; \sum_{i \,\in\, I_k} (p_i - q_i)^2,\ and\ 
S_k^2(\mathbf{p}, \mathbf{q}) \;=\; \sum_{j=1}^{k} d_j^2(\mathbf{p}, \mathbf{q}) , 
\]
respectively. If the current best candidate in the top-$K$ search has a distance threshold $\tau^2$, then as soon as
$S_k^2(\mathbf{p}, \mathbf{q}) \;>\; \tau^2$,
we know that adding further (non-negative) contributions from machines $k+1$ through $M$ will only increase $S_k^2(\mathbf{p}, \mathbf{q})$. Consequently, no matter what the remaining dimensions contribute, $\mathbf{q}$ cannot enter the top-$K$ set. We thus \emph{prune} $\mathbf{q}$ and skip the remaining machines.
This gives us the possibility to further reduce the computational overhead. This approach significantly reduces computational overhead.

\begin{figure}
    \centering
    \begin{subfigure}{0.48\linewidth}
        \includegraphics[width=\linewidth,height=0.64\linewidth]{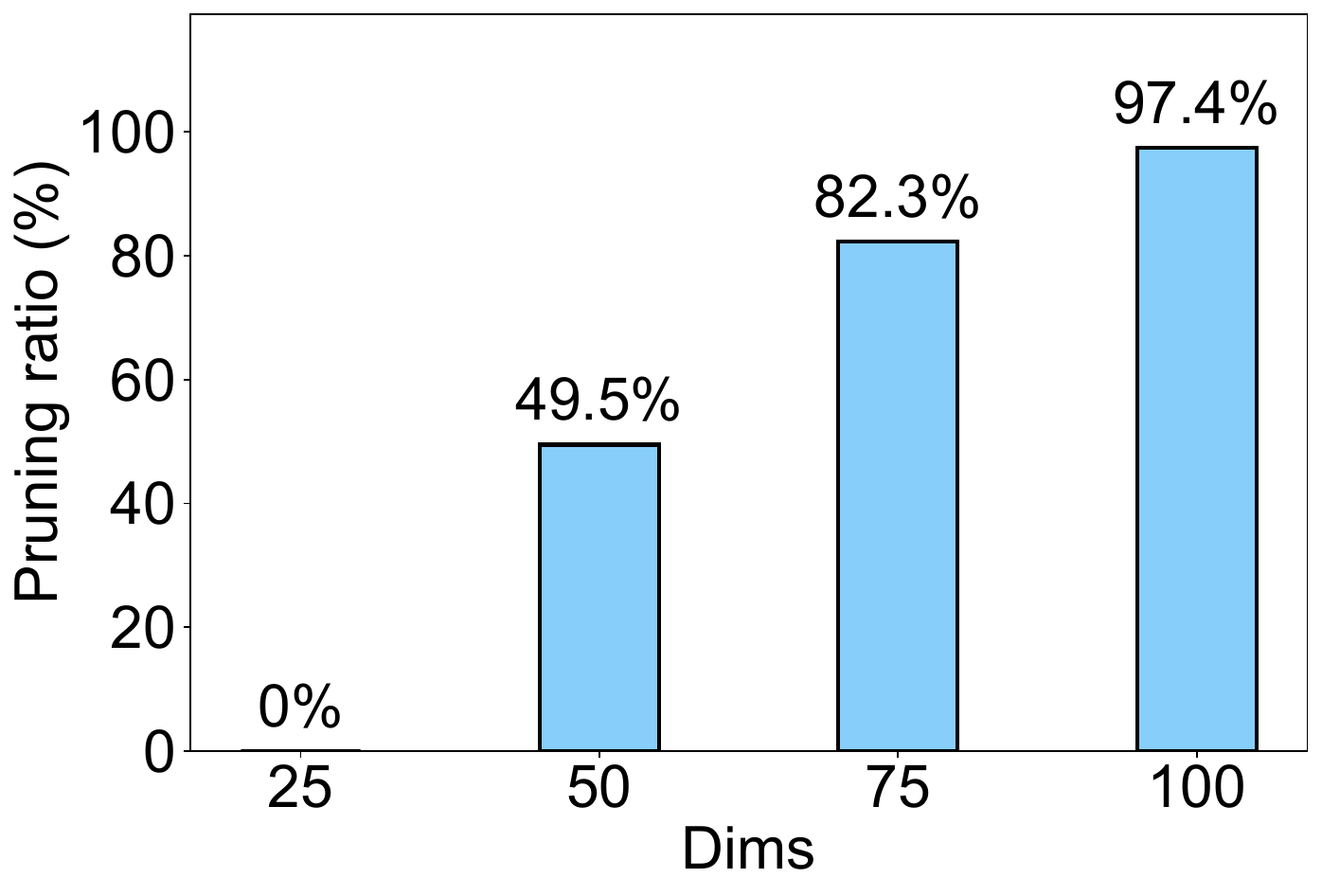}
        \caption{Pruning ratio by dimensions}
        \label{fig:base_sub2}
    \end{subfigure}
    \hfill
    \begin{subfigure}{0.48\linewidth}
        \includegraphics[width=\linewidth,height=0.64\linewidth]{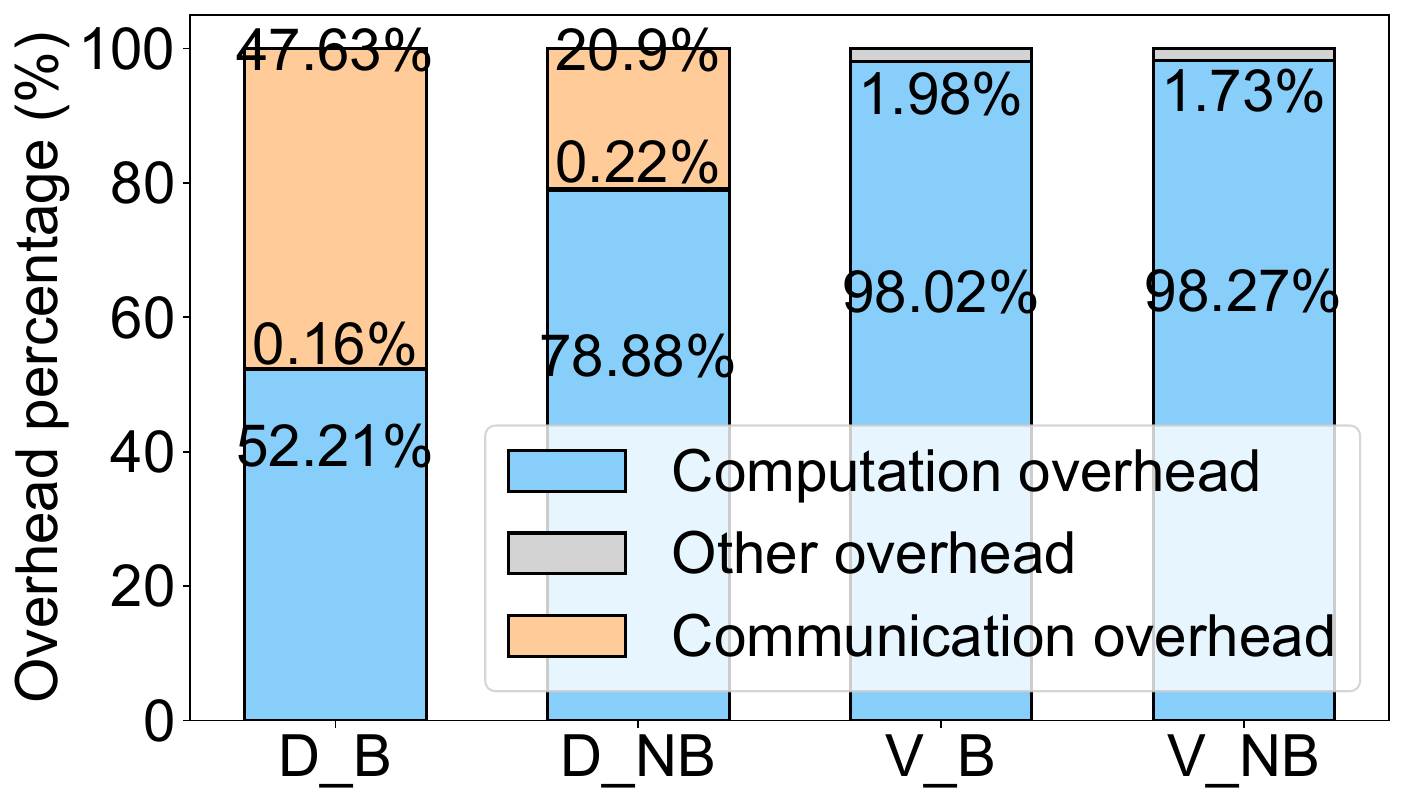}
        \caption{Time overhead breakdown}
        \label{fig:base_sub1}
    \end{subfigure}
        \vspace{-0.13in}

    \caption{Pruning ratio and time overhead breakdown.}
    \label{fig:microbenchmark}
        \vspace{-0.1in}
    \vspace{-0.1in}
        \vspace{-0.02in}

\end{figure}

\textbf{Empirical results.}
To illustrate the effectiveness of dimension-level pruning, we present empirical results in Figure~\ref{fig:microbenchmark}(a), which showing the pruning ratio (the fraction of distance calculations skipped) for a setting with four machines, each responsible for one-quarter of the vector dimensions (i.e., $[1,\tfrac{d}{4}]$, $[\tfrac{d}{4}+1,\tfrac{d}{2}]$, etc.). Notably:
\begin{itemize}[leftmargin=1.5em]
    \item By the time the second machine completes, nearly 50\% of the candidate vectors are pruned.
    \item For the third and fourth machines, the pruning ratio exceeds 80\%, reaching up to 97.4\% in our experiment.
\end{itemize}
These results confirm that dimension-level pruning significantly reduces computational overhead and network transfers when partial distances can rule out most candidates early, further illustrating the benefits of distributing vectors by dimension.

\textbf{Motivation 2: Cost breakdown and skewed loads effects in distributed vector databases.}

\textbf{Cost breakdown.} We evaluate two commonly used partitioning strategies: dimension-based partitioning (D) and vector-based partitioning (V) under blocking (B) and non-blocking (NB) communication modes. We run nearest neighbor searches on the Sift1M dataset using a five-node cluster (one client node and four worker nodes), each with 56 vCPUs/threads and interconnected via 100 Gb/s links. The breakdown of computation (blue), communication (orange), and other overhead (gray) is presented in Figure~\ref{fig:microbenchmark}~(b).



Under both blocking (B) and non-blocking (NB) communication modes, vector-based partitioning (V) consistently incurs lower communication overhead than dimension-based partitioning (D). On average, the overall communication time for V is reduced by about 66\% relative to D, which aligns with our hypothesis that dimension-based partitioning requires additional data exchange steps to assemble partial distance results.

\textbf{Stability.} We observe a stark contrast in query bandwidth stability under balanced and imbalanced workloads between the two partitioning approaches. Specifically, dimension-based partitioning delivers a consistently stable query bandwidth, regardless of whether the load is evenly distributed across nodes. 
In contrast, vector-based partitioning achieves 30\% higher bandwidth than dimension-based partitioning under balanced loads, but its performance degrades substantially under load imbalance.
With skewed workloads in a five-node setup, vector-based partitioning can drop to as low as 26\% of the cluster-based approach’s bandwidth, underscoring the trade-off between maximizing throughput under balanced conditions and maintaining stability when workloads vary.


\subsection{Challenges}
\label{subsec:challenges}
Distributed vector databases promise scalable and efficient approximate nearest neighbor search (ANNS), but three key challenges must be addressed to realize their potential.
 


\textbf{Challenge 1: Combining partitioning strategies in ANNS.}
A key challenge is designing an indexing scheme that integrates both vector-based and dimension-based partitioning effectively. 
Vector-based partitioning reduces communication overhead, while dimension-based partitioning balances workloads.
However, merging these strategies requires adaptive mechanisms to exploit their strengths under varying query distributions.
Achieving high throughput and stable performance across diverse workloads requires dynamic partitioning policies and runtime optimizations.

\textbf{Challenge 2: Efficient query distribution in ANNS.}
Even with robust partitioning, queries must be efficiently distributed across machines to minimize communication and computation overhead while ensuring load balancing. Merely splitting queries among nodes does not guarantee the avoidance of bottlenecks. Consequently, innovative query assignment and workload distribution techniques are necessary to allocate queries in a manner that reduces network traffic, prevents node saturation, and maintains low latency. These mechanisms should also adapt to shifting workloads and handle skewed query distributions seamlessly.

\textbf{Challenge 3: Multi-machine pruning in high dimension.}
Finally, multi-machine environments introduce opportunities for inter-machine pruning, where partial results can eliminate the need for further computations on certain nodes. Yet, coordinating partial results in a way that swiftly terminates redundant processes without introducing excessive overhead remains non-trivial. Designing lightweight pipelines to propagate pruning decisions efficiently and synchronize nodes with minimal communication overhead is paramount for achieving robust query efficiency in large-scale ANNS systems.

\section{System Design}
\label{sec:system}

\subsection{Overview}
We introduce \sysname{}, a distributed ANNS engine that can address the limitations of existing distributed partitioning strategies, such as difficulty maintaining balanced loads and achieving high throughput.
\sysname{} combines multi-granularity partition with dimension-level pruning. 
Like other vector search engines~\cite{NSG,facebook2020a_SOSP_17}, \sysname{} independently indexes and performs approximate searches on large-scale vector collections.
Unlike prior works~\cite{Spann,diskann,NSDI2023}, \sysname{} adapts its partitioning strategies to dynamic query workloads and uses dimension-based early-stop mechanisms to prune costly distance computations.

\textbf{Solution to challenges.}
\sysname{} tackles the three major challenges described in Section~\ref{subsec:challenges} through the following key ideas.

\paragraph{1) Adaptive partitioning strategy for optimal throughput and stability.} To combine vector-based and dimension-based partitioning, \sysname{} adopts a dynamic approach that uses a cost model to determine the most efficient partition strategy based on the current load distribution. The system balances vector-based partitioning, which reduces communication overhead, with dimension-based partitioning, which ensures workload balance. This adaptive scheme enhances throughput and stability across diverse query workloads, optimizing the benefits of both approaches.

\paragraph{2) Efficient query distribution with load-aware routing.} For efficient query distribution, \sysname{} implements a load-aware routing mechanism that minimizes communication overhead while ensuring balanced query processing. The system adjusts query assignments based on real-time workload analysis, ensuring nodes are neither overwhelmed nor underutilized. 
By distributing queries to minimize commnuication overhead and prevent node saturation, \sysname{} ensures that query execution is both fast and balanced, even under skewed query distributions.

\paragraph{3) Multi-machine pruning with lightweight pipelines.} To tackle the challenge of multi-machine pruning, \sysname{} introduces a lightweight inter-machine pruning mechanism that propagates partial results and prunes irrelevant computations across nodes. By synchronizing nodes with minimal communication overhead, the system reduces redundant computation, accelerating query execution and leveraging the parallelism of distributed environments to minimize unnecessary overhead in large-scale ANNS systems.

\begin{figure}
    \centering
    \includegraphics[width=\linewidth,height=0.35\linewidth]{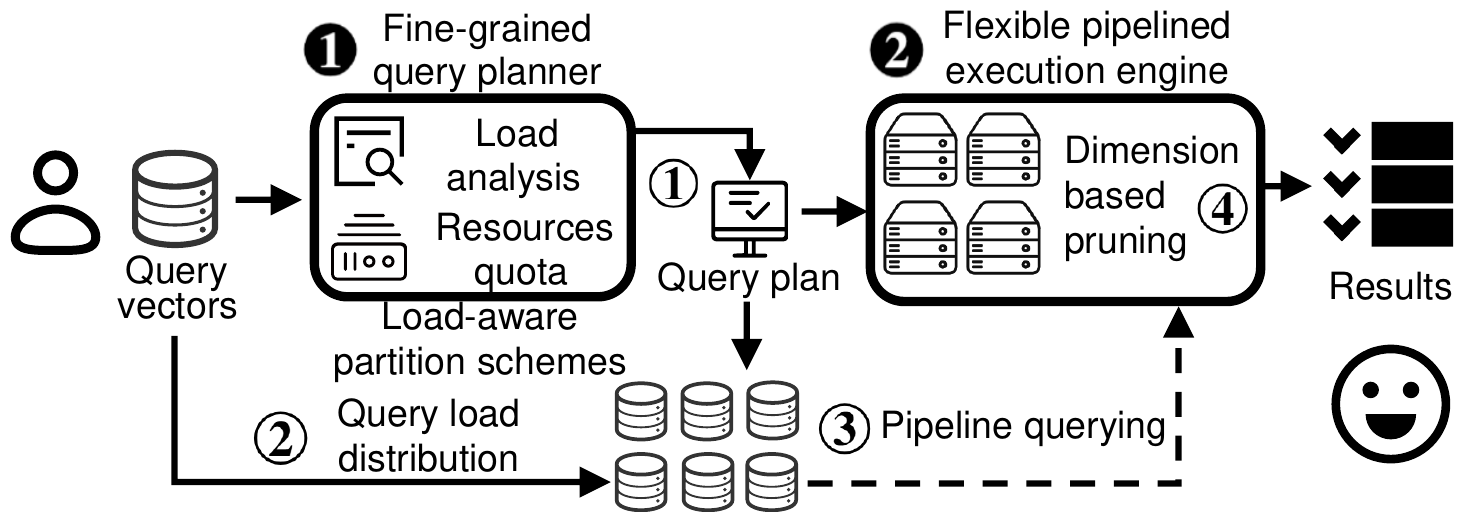}

    \caption{\sysname{}'s query process.}
    \label{fig:harmonydesign}

\vspace{-.23in}

\end{figure}

\textbf{Models.}
\sysname{} is designed with two key modules: 
\textbf{\tikz[baseline=(char.base)] \node[draw, circle, fill=black, text=white, inner sep=0.5pt, scale=0.7] (char) {1};}  fine-grained query planner (detailed in Section~\ref{subsec:fine-grained})
and 
\textbf{\tikz[baseline=(char.base)] \node[draw, circle, fill=black, text=white, inner sep=0.5pt, scale=0.7] (char) {2};}
flexible pipelined execution engine (detailed in Section~\ref{subsec:pipelined_engine}). 
Figure~\ref{fig:harmonydesign} illustrates the overall workflow and how each module addresses the primary challenges.

\begin{itemize}[leftmargin=1.25em]
    \item \textbf{\tikz[baseline=(char.base)] \node[draw, circle, fill=white, text=black, inner sep=0.5pt, scale=0.7] (char) {1}; Load-aware query plan} and \textbf{\tikz[baseline=(char.base)] \node[draw, circle, fill=white, text=black, inner sep=0.5pt, scale=0.7] (char) {2}; Query load distribution}. These components analyze resource usage and query skewness, leveraging a cost model to select between dimension-based, vector-based, or hybrid partitioning. Building on the chosen partition strategy, \sysname{} distributes query vectors across nodes according to the anticipated workload, ensuring balanced resource utilization.

   \item \textbf{\tikz[baseline=(char.base)] \node[draw, circle, fill=white, text=black, inner sep=0.5pt, scale=0.7] (char) {3}; Pipeline querying} and \textbf{\tikz[baseline=(char.base)] \node[draw, circle, fill=white, text=black, inner sep=0.5pt, scale=0.7] (char) {4};  Dimension-based pruning}. To avoid parallel processing of all vector dimensions at each node, these components adopt a pipeline-based flow. Partial results are passed sequentially along nodes to avoid duplicative work on candidates with pruning potential. As partial results accumulate, vectors exceeding the pruning threshold are discarded before reaching other nodes. 
\end{itemize}

These components work together to adaptively partition data based on the current workload and exploit dimension-level pruning, ensuring robust performance and reduced query latencies across diverse high-dimensional scenarios.

\textbf{Novelties.} \sysname{} offers three key innovations:

\begin{itemize}[noitemsep,nolistsep,leftmargin=1.25em]
    \item \textbf{Adaptive index construction.} 
    \sysname{} adopts an adaptive indexing approach that dynamically adjusts the index structure based on workload characteristics. This ensures efficient handling of varying query types and workloads, maintaining high performance across scenarios.

    \item \textbf{Load-aware partitioning for balanced distribution.} 
    By integrating dimension-based and vector-based partitioning, \sysname{} ensures that vectors are distributed across nodes to minimize load imbalance. This ensures stable performance under skewed loads while preserving high throughput for balanced workloads.

    \item \textbf{Dimension-level pruning.} 
    \sysname{} incorporates dimension-level pruning in its query processing pipeline. By executing partial distance calculations across nodes, vectors that cannot exceed the current best threshold are pruned early, reducing unnecessary computation and communication, especially for high-dimensional queries.
\end{itemize}


\subsection{Fine-grained Query Planner}
\label{subsec:fine-grained}
\begin{table}[t]
\centering
\caption{Notations for Our Cost Model}
\label{tab:notations}
\begin{tabular}{@{}ll@{}}
\toprule
\textbf{Symbol}                                   & \textbf{Meaning}                                                                 \\ 
\midrule
$\pi$                                             & Partition plan                                                                     \\
$\mathcal{B}_{\text{dim}}(\pi), \;\mathcal{B}_{\text{vec}}(\pi)$ 
                                                 & Dimension- and vector-based blocks in $\pi$ \\
$b,\,s$                                           & Single block $b\in\mathcal{B}_{\text{dim}}(\pi)$,  
                                                   $s\in\mathcal{B}_{\text{vec}}(\pi)$                                               \\
$
c_{\mathrm{comp}}^{\mathrm{vec}}(b,q),\,
c_{\mathrm{comm}}^{\mathrm{dim}}(s,q), $                                                                                                        & Computation/communication costs for  \\ 
& query $q$ when processed in  $b$ or $s$                                         \\[4pt]
$\text{Load}(n, \pi)$                             & Total load on node $n$ under plan $\pi$                                           \\
$\mathcal{I}(\pi)$                                & Imbalance factor                         \\
$\alpha$                                          & Weight for imbalance in overall cost                                              \\
\bottomrule
\end{tabular}
\end{table}

To achieve optimal partitioning and query strategy, we introduce a cost model that evaluates computational and communication overhead during execution. Table~\ref{tab:notations} shows the key variables and metrics used in the cost model, including partition definitions, query processing costs, and imbalance factors. Like traditional databases, computational and transmission overheads can be efficiently estimated during the initial query setup. 
For instance, lightweight metrics such as the number of cluster centers in distance calculations or the dimensionality of query vectors allow the cost model to be computed with minimal overhead.

\subsubsection{Cost model.}\label{subsubsce:cost_mod} Let
$  \mathcal{Q} = \{\,q_1,\,q_2,\dots,\,q_{|\mathcal{Q}|}\}$
 the set of query vectors.

\textbf{Query cost.}
For each query $q \in \mathcal{Q}$, we sum two categories of partitioned data: 
(i) dimension-based partitions and (ii) vector-based partitions. The per-query cost is:
\[
\begin{aligned}
  \mathcal{C}_{q}(\pi)
  &\;=\;
  \underbrace{
  \sum_{b \,\in\, \mathcal{B}_{\text{dim}}(\pi)}
  \Bigl[
    c_{\mathrm{comp}}^{\mathrm{dim}}(b,q)
    \;+\;
    c_{\mathrm{comm}}^{\mathrm{dim}}(b,q)
  \Bigr]
  }_{\text{Dimension-based component}} \\
  &\;+\;
  \underbrace{
  \sum_{s \,\in\, \mathcal{B}_{\text{vec}}(\pi)}
  \Bigl[
    c_{\mathrm{comp}}^{\mathrm{vec}}(s,q)
    \;+\;
    c_{\mathrm{comm}}^{\mathrm{vec}}(s,q)
  \Bigr]
  }_{\text{Vector-based component}}
  ,
\end{aligned}
\]
where $\mathcal{B}_{\text{dim}}(\pi,n)$ denotes the dimension-based blocks
stored on node $n$, and $\mathcal{B}_{\text{vec}}(\pi,n)$ the vector-based shards on node $n$.

\textbf{Imbalance factor.}
We define $\text{Load}(n,\pi)$ to measure the total work node $n$ performs:
\[
  \text{Load}(n,\pi)
  =
  \sum_{q\in \mathcal{Q}}
  \bigl[
    \sum_{b\in\mathcal{B}_{\text{dim}}(\pi,n)}
      c_{\mathrm{comp}}^{\mathrm{dim}}(b,q)
    +
    \sum_{s\in\mathcal{B}_{\text{vec}}(\pi,n)}
      c_{\mathrm{comp}}^{\mathrm{vec}}(s,q)
  \bigr],
\]

The \emph{imbalance factor} $\mathcal{I}(\pi)$ is defined via the standard deviation:
\[
  \mathcal{I}(\pi)
  \;=\;
  \sqrt{
    \frac{1}{N}
    \sum_{n=1}^{N}
    \bigl[
      \text{Load}(n,\pi) 
      - 
      \overline{L}
    \bigr]^2
  },
  \;
  \text{where}
  \;
  \overline{L}
  =
  \frac{1}{N}\sum_{n=1}^N
  \text{Load}(n,\pi).
\]
A larger imbalance implies that certain nodes may overload.

\textbf{Overall cost function.}
We combine the per-query costs and imbalance term into a single objective:
\[
  \mathcal{C}\bigl(\pi,\mathcal{Q}\bigr)
  \;=\;
  \sum_{q \,\in\, \mathcal{Q}}
  \mathcal{C}_{q}(\pi)
  \;+\;
  \alpha\,\cdot\,\mathcal{I}(\pi).
\]
Here, $\alpha$ is a user-defined weight emphasizing how severely we penalize
skew vs.\ local comp/comm efficiency.

\textbf{Example application.} 
To demonstrate how the cost model guides partitioning, consider a cluster of four worker nodes and one client node. Assume the following query characteristics: $c_{\mathrm{comp}}^{\mathrm{dim}}(b,q) = 20$ ms; $c_{\mathrm{comm}}^{\mathrm{dim}}(b,q) = 30$ ms; $c_{\mathrm{comp}}^{\mathrm{vec}}(s,q) = 15$ ms; $c_{\mathrm{comm}}^{\mathrm{vec}}(s,q) = 1$ ms.

Initially, the workload is divided such that the dataset is partitioned into $3$ dimension-based blocks and $2$ vector-based shards. The imbalance factor $\mathcal{I}(\pi)$ contributes about $5\%$ of the overall delay due to slight workload variations across the nodes. However, communication costs dominate the overall query latency, the current partitioning plan results in suboptimal performance.

In this case, the cost model suggests adjusting the granularity of the partitions. 
Specifically, reducing dimension-based blocks from 3 to 2 and increasing vector-based shards from two to three would better balance the workload.
This adjustment minimizes communication overhead by leveraging vector-based partition's lower cost while maintaining a manageable computation load across the nodes.

\begin{figure}
    \centering
     \includegraphics[width=\linewidth,height=0.52\linewidth]{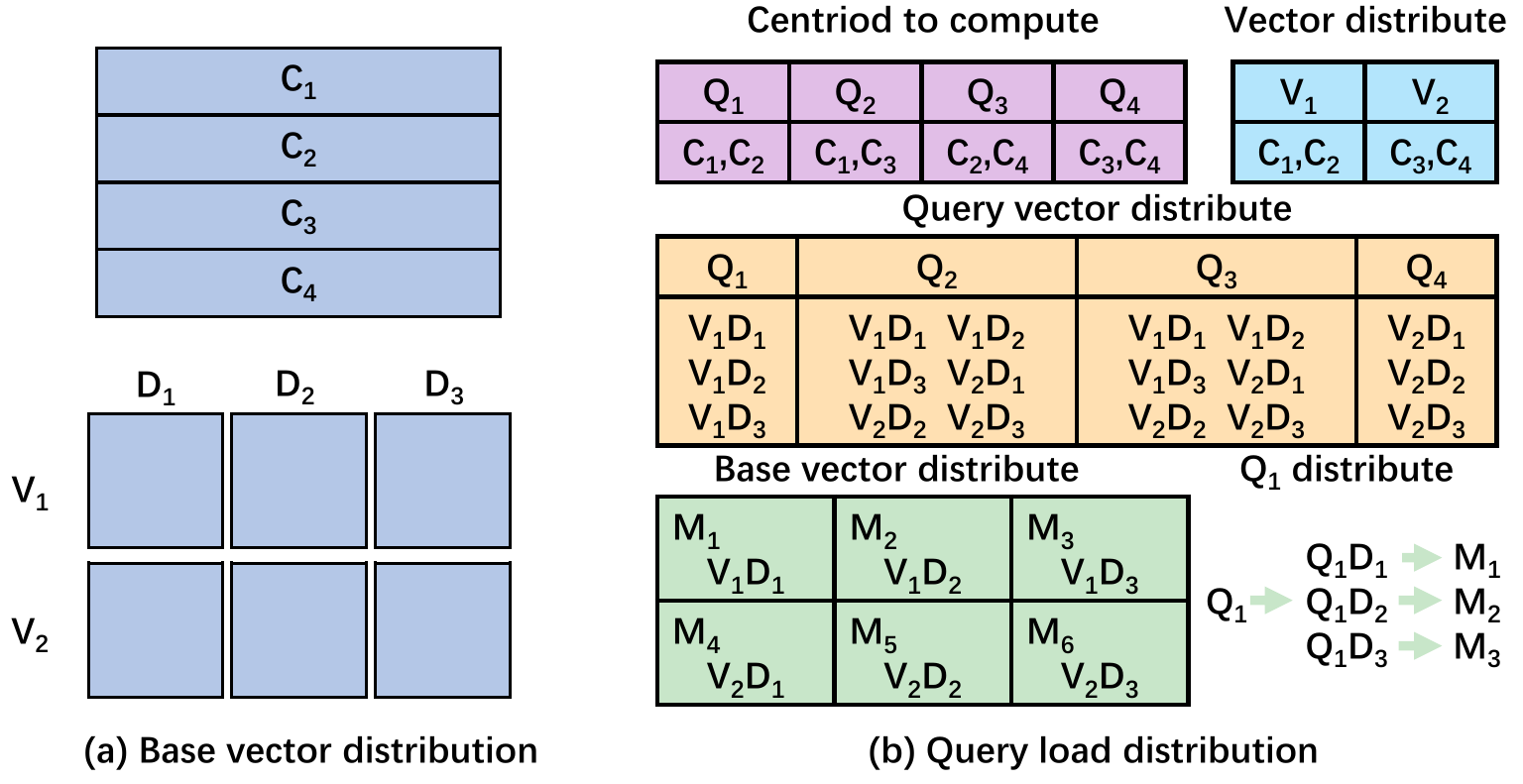}

    \caption{\sysname{}'s query distribution. C${_i}$ denotes the id of the cluster, Q${_i}$ denotes the query block i, V${_i}$ denotes the ith block according to vector-based partition, D${_i}$ denotes the ith block according to dimension-based partition, M${_i}$ denotes the ith machine.}
    \label{fig:harmonydistribute}
\vspace{-0.15in}
\end{figure}

\subsubsection{Query load distribution.}
Figure~\ref{fig:harmonydistribute} illustrates how queries are distributed.
First, after the cost model analyzes the workload and finalizes a partition plan $\pi$, 
we obtain a grid of dimension splits $\mathcal{B}_{\text{dim}}(\pi)$ 
and vector splits $\mathcal{B}_{\text{vec}}(\pi)$. 
As shown in Figure~\ref{fig:harmonydistribute}~(a), 
dimension partitions $\{\text{D${_1}$}, \text{D${_2}$}, \text{D${_3}$}\}$ 
and vector partitions $\{\text{V${_1}$}, \text{V${_2}$}\}$, 
result in blocks $\{\text{V${_1}$D${_1}$}, \text{V${_1}$D${_2}$},\dots\}$.
Each machine in the cluster ($M_1$--$M_6$) holds one of these blocks completing the base vector distribution.

Second, after distributing the base vector, \sysname{} completes the query load distribution when the query vector is received, as shown in Figure~\ref{fig:harmonydistribute}~(b):

\begin{enumerate}
    \item \textbf{Identify cluster centroids (purple table).}
    Like many cluster-based ANNS engines, \sysname{} first retrieves the relevant centroids on the client side to map each query $q_i$ to the associated portion of the base vector.

    \item \textbf{Map queries to vector-based blocks (blue table).}
    After retrieving the centroids, \sysname{} determines which vector partitions
    $\mathcal{B}_{\text{vec}}(\pi)$ each query should visit. For example, in Figure~\ref{fig:harmonydistribute}(b), $Q1$ is mapped to $\text{V1}$.

    \item \textbf{Combine with dimension splits and map to machines (orange and green tables).}
    Once \sysname{} determines the vector partitions $\mathcal{B}_{\text{vec}}(\pi)$ for each query, 
    it splits the queries based on dimension partitions $\mathcal{B}_{\text{dim}}(\pi)$. 
    For example, $Q1$ mapped to $\text{V1}$ will be split into $Q1D1$, $Q1D2$, and $Q1D3$ 
    corresponding to dimension partitions $\{\text{D1}, \text{D2}, \text{D3}\}$. 
    These query chunks are routed to corresponding machines for partial distance computations. 
    As shown in Figure~\ref{fig:harmonydistribute}(b), 
    blocks like $\text{V1D1}$, $\text{V1D2}$, and $\text{V1D3}$ are assigned to $M1$, $M2$, and $M3$. 
    This ensures $Q1D1$, $Q1D2$, and $Q1D3$ are processed in parallel on separate machines. 
    By evenly distributing these workloads, \sysname{} avoids bottlenecks and fully utilizes cluster resources.
\end{enumerate}

\textbf{Advantages.}
Because the blocks in $\mathcal{B}_{\text{dim}}(\pi)\times\mathcal{B}_{\text{vec}}(\pi)$ 
are evenly distributed among machines, the four queries fully utilize all six nodes. 
No machine handles a disproportionately large portion of the dataset, resulting in balanced load and higher throughput.

\textbf{Time complexity.} The time complexity of \sysname{}’s query distribution is determined by two factors: centroid assignment and query distribution across machines. 

1. \emph{Centroid Assignment (Computation overhead).} For $Q$ query vectors with $D$ dimensions and $NC$ base centroids, finding the closest centroids costs $O(Q \cdot NC \cdot D^2)$, a standard operation in most ANNS engines without additional overhead specific to \sysname{}. 

2. \emph{Query Distribution (Communication overhead).} After centroid assignment, each query vector is distributed across machines for partial distance computations. Assume a query vector needs to interact with $V$ machines. With \sysname{}, this number increases to $V \cdot B$, where $B$ is the number of dimension-based splits $\mathcal{B}_{\text{dim}}(\pi)$.
However, the total data sent does not change, as the size of each split is reduced to $Q \cdot D / B$. As a result, while the query might involve more communication, the total communication cost remains the same.

Overall, compared to traditional partitioning schemes, \sysname{} does not add any communication or computation overhead. 
By keeping the total data transferred unchanged and leveraging efficient centroid assignment, Harmony achieves comparable performance and fine-grained load balancing.
This claim is further validated experimentally in Section~\ref{subsce:overall_performenca}.

\textbf{Space Complexity.} \sysname{}’s space requirements are minimal as each base vector is stored on one machine, eliminating redundancy. The query set is typically small compared to base vectors, so storing and routing query splits incur negligible space costs. These space requirements ensure Harmony can scale effectively without significant memory consumption. We further validate this claim in Section~\ref{subsubsec:index_memory}.

\subsection{Flexible Pipelined Execution Engine}
\label{subsec:pipelined_engine}

\sysname{} leverages multiple machines to handle parts of a query’s vectors or dimensions, enabling distributed pruning. 
However, this approach presents a challenge mentioned in Section~\ref{sec:intro}: when each machine computes partial distances in isolation, it loses the ability to prune candidates based on results from other machines.

\textbf{Solution.} To resolve this, \sysname{} adopts a pipelined execution scheme, ensuring once a candidate vector or part of its dimension is deemed unpromising, no further machines need to process it.

\begin{algorithm}[t]
\footnotesize
\DontPrintSemicolon
\caption{\textsc{FlexiblePipelinedPruning}}
\label{alg:query_pipeline}

\SetKwFunction{FPrewarm}{PrewarmHeap}
\SetKwFunction{FVectorPipe}{VectorPipeline}
\SetKwFunction{FDimensionPipe}{DimensionPipeline}
\SetKwFunction{FUpdatePrune}{UpdatePruning}
\SetKwFunction{FComputeDist}{ComputeDistance}

\SetKwProg{Fn}{Function}{:}{\KwRet}

\Fn{\FPrewarm(QuerySet Q, int K)}{
    \tcp{Compute partial distances for a subset of queries to build an initial max-heap.}
    \ForEach{q in Q\_subset}{
        dist $\leftarrow$ \FComputeDist(q, randomVectors)\;
        heap\_insert(q, dist, K)\;
    }
    \textbf{return} heap\;
}

\Fn{\FDimensionPipe(q, DSet)}{
    \tcp{Process each dimension block in DSet sequentially.}
    \ForEach{d in DSet}{
        partialDist $\leftarrow$ \FComputeDist(q, d)\;
        \If{partialDist > q.currentThreshold}{
            prune(q)\;
            \textbf{return}\;  
        }
        \FUpdatePrune(q, partialDist)\;
    }
}

\Fn{\FVectorPipe(QueryBatch Qv, DSet)}{
    \tcp{Process a batch of queries for a specific vector partition.}
    \ForEach{q in Qv}{
        \FDimensionPipe(q, DSet)\;
        \If{q.pruned == true}{
            \textbf{continue}\;
        }
        \FUpdatePrune(q, finalDist)\;
    }
}

\Fn{\textsc{QueryPipeline}(QuerySet Q, VSetList VList, DSetList DList, int K)}{
    \tcp{Stage 0: Prewarming}
    heap $\leftarrow$ \FPrewarm(Q, K)\;
    
    \tcp{Stage I: Vector-level pipeline}
    \ForEach{vPart in VList}{
        Qv $\leftarrow$ filterQueries(Q, vPart)\;
        \FVectorPipe(Qv, DList)\;
    }
}
\end{algorithm}

Algorithm~\ref{alg:query_pipeline} illustrates how \sysname{} orchestrates vector-level and dimension-level distance computations in a pipeline. We first describe the algorithmic structure, then highlight the key lines.
\begin{enumerate}[leftmargin=*]
\item \textbf{Prewarm stage (Lines~1-5).}    
    \(\FPrewarm(Q, K)\) calculates initial distances for a subset of queries stored in the client node, including the centroid and a few other vectors. This populates a max-heap of size \(K\), providing an early pruning threshold.

\item \textbf{Vector pipeline (Lines~13--18).}  
\(\FVectorPipe(Qv, DList)\) iterates over each vector partition \(\mathcal{V}\) in \(\mathit{VList}\).  
Within a single vector-based search, it invokes the dimension pipeline (Line~15) to handle partial distance computations. After processing each vector partition, the pipeline updates pruning thresholds and discards queries no longer qualifying for the top-$K$ set. 
This ensures each machine specializes in searching a particular vector partition, ensuring parallelism while respecting the global pruning condition.

\item \textbf{Dimension pipeline (Lines~6--12).}  
\(\FDimensionPipe(q, DSet)\) scans each dimension block in $\mathcal{B}_{\text{dim}}(\pi)$ sequentially. For each block, the dimension pipeline computes a partial distance and compares it to the current pruning threshold. If the distance exceeds the threshold, the query is pruned immediately, skipping further dimensions. Otherwise, the pipeline updates the threshold to reflect the partial result, potentially lowering it for subsequent blocks. This fine-grained control allows unpromising queries to exit early, minimizing redundant computations across the cluster.

\end{enumerate}

\textbf{Complexity.} We analyze \sysname{}'s complexity under balanced load. Let $Q$ denote the number of query vectors, $NB$ the number of base vectors (data points), and $D$ the dimensionality of each vector, with distance computations costing up to $O(D^2)$ in the naive case. Additionally, let $n_{\text{list}}$ denote the total number of clusters (lists) and  $n_{\text{probe}}$ the number of probed clusters per query. These parameters define the computational and communication load across vector database.

\paragraph{Time Complexity.}
In a naive (centralized) scenario, one might compute $O(Q \cdot NB \cdot n_{\text{probe}} /n_{\text{list}} \cdot D^2)$ total operations. However, we typically distribute work across $\mathcal{B}_{\text{vec}}(\pi)\,\times\,\mathcal{B}_{\text{dim}}(\pi)$ machines. Under balanced load, each machine handles:
\[
\frac{
  Q \,\times\, (n_{\text{probe}} / n_{\text{list}}) \,\times\, NB \,\times\, D^2
}{
  \mathcal{B}_{\text{vec}}(\pi) \,\times\, \mathcal{B}_{\text{dim}}(\pi)
}.
\]

Thus, the final per-machine cost scales as 
$O\!\bigl(\tfrac{Q \cdot n_{\text{list}} \cdot NB \cdot D^2}{n_{\text{probe}} \,\mathcal{B}_{\text{vec}}(\pi)\,\mathcal{B}_{\text{dim}}(\pi)}\bigr)$. The degree of computational reduction is proportional to the number of machines.

\paragraph{Communication Overhead.}
On average, each query interacts with 
\(
  n_{\text{probe}} \,\times\, \tfrac{\mathcal{B}_{\text{vec}}(\pi)}{n_{\text{list}}}
\)
vector partitions. Each partition has $\mathcal{B}_{\text{dim}}(\pi)$ dimension blocks. The number of communication required is $\mathcal{B}_{\text{dim}}(\pi)$ times greater, but the amount of transfers also becomes $1/\mathcal{B}_{\text{dim}}(\pi)$
Consequently, the communication for query distribution grows proportionally to:
\[
\left(\tfrac{NB}{\mathcal{B}_{\text{vec}}(\pi)}\right)
\;\times\;
\Bigl(
  n_{\text{probe}} \,\times\, 
  \tfrac{\mathcal{B}_{\text{vec}}(\pi)}{n_{\text{list}}}
\Bigr)
\;\times\;
\tfrac{D}{\mathcal{B}_{\text{dim}}(\pi)}
\;\times\;
\mathcal{B}_{\text{dim}}(\pi)
,
\]
thus, the communication overhead in \sysname{} remains identical to a purely vector-based scheme, as the finer partitioning of dimensions does not increase the total amount of transmitted data—it only reorganizes the transmission into smaller, parallelizable chunks.

\paragraph{Space Complexity.}
Each of the $NB$ base vectors is split among $\mathcal{B}_{\text{vec}}(\pi)$ (vector) $\times \mathcal{B}_{\text{dim}}(\pi)$ (dimension) blocks. Thus, each block stores roughly 
\[
\frac{NB \,\times\, D}{\mathcal{B}_{\text{vec}}(\pi)\,\times\,\mathcal{B}_{\text{dim}}(\pi)}
\]
elements, giving total storage of $NB \cdot D$ (no duplication). Intermediate query results are small compared to the raw dataset and can be ignored in practice. Consequently, the space cost remains $O(NB \cdot D)$, matching standard distributed systems without extra replication.

\begin{figure}
    \centering
    \includegraphics[width=\linewidth,height=0.63\linewidth]{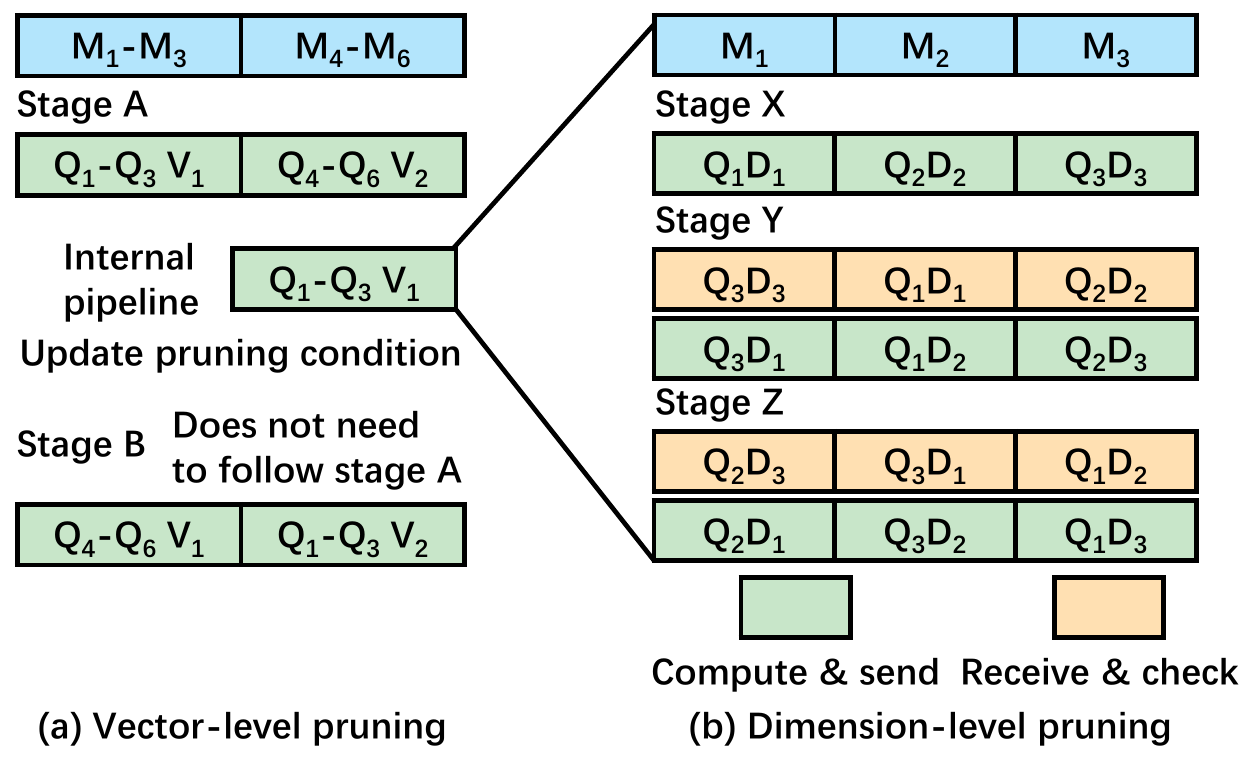}

    \caption{\sysname{}'s pipeline pruning and querying.}
    \label{fig:harmonypipline}
\vspace{-0.2in}
\end{figure}

\textbf{Example.} Figure~\ref{fig:harmonypipline} shows an example of pipeline and pruning strategy for Harmony’s vector and dimension granularity. Assuming query vectors $Q1-Q6$ need to compute distances with base vector blocks $V1, V2$.

\emph{Vector-level pipeline and pruning.}
In the vector-level pipeline (Figure~\ref{fig:harmonypipline}~(a)), queries are processed in batches corresponding to vector partitions (\(\mathcal{B}_{\text{vec}}(\pi)\)). For example, during Stage A, queries \(Q1\)--\(Q3\) are assigned to vector partition \(V1\), while \(Q4\)--\(Q6\) are assigned to \(V2\). 
Each batch completes its initial distance calculations and updates the global pruning condition by refining the max-heap threshold. This ensures that subsequent query batches (\(Q4\)--\(Q6\) and \(V1\) in Stage B) benefit from a tighter pruning threshold, reducing unnecessary computations. 
Iteratively refining the heap after processing each batch, the vector-level pipeline achieves efficient pruning while maintaining parallelism across vector partitions.

\paragraph{Dimension-Level Pipeline.}
In the dimension-level pipeline (Figure~\ref{fig:harmonypipline}(b)), each vector's dimensions are distributed across machines, ensuring no multiple dimensions are processed simultaneously.
This design guarantees dimension-wise computations proceed in a pipeline fashion, enabling effective pruning.

In Stage X, each machine begins processing the first assigned dimension block of the query. For instance, \(Q1\) begins with \(D1\) on \(M1\), \(Q2\) starts with \(D2\) on \(M2\), and \(Q3\) with \(D3\) on \(M3\). After computing the partial distance for the first block, intermediate results are accumulated and compared against the global pruning threshold. If the accumulated distance exceeds the threshold, the query is pruned early, preventing further computations for the remaining dimensions.

In Stage Y, machines process the next dimension blocks while respecting the updated pruning conditions. For example, \(Q1D2\) is sent to \(M2\), \(Q2D3\) to \(M3\), and \(Q3D1\) to \(M1\). During this stage, the partial results computed in Stage X are reused and incremented with the newly calculated distances from the current dimension block. The pruning threshold is checked again, ensuring no unnecessary computation for unpromising queries.

Between stages, workers transmit partial results, which are lightweight compared to the original data. Furthermore, by ensuring that different dimensions of the same query vector are processed in non-consecutive stages, such as introducing additional queries (\(Q7\)--\(Q9\)), the partial result transmission does not block the critical computation path. Each stage proceeds independently without waiting for the previous stage, significantly reducing communication overhead.


This pipeline ensures that, at each stage, dimensions for a given query are separated across machines, eliminating overlap in computation. By leveraging the distribution of dimensions across nodes, Harmony avoids synchronization overhead typically required in dimension-partition implementations. This approach maximizes pruning while minimizing inter-machine communication and redundant calculations.

\textbf{Advantages.}
By combining vector-level and dimension-level pipelines, \sysname{} achieves coarse-grained and fine-grained pruning in a single framework. 
Specifically, the vector-level pipeline updates a global threshold after each batch, preventing unnecessary computations for subsequent ones.
Meanwhile, the dimension-level pipeline processes dimension blocks in sequence, discarding queries whose partial distance exceeds the threshold.
This dual-stage approach exploits the hierarchical distribution of data and queries, minimizing computation and communication overhead.
Unlike previous methods that restrict dimension-based pruning to a single node~\cite{ADSampling}, \sysname{} distributes dimensions across machines, eliminating interrupt-driven pruning checks and reducing synchronization overhead. This design also facilitates integration with hardware accelerations (e.g., SIMD), as the dimension-level pipeline decouples partial result transmissions from compute operations. Queries are pruned in real time, the workload is balanced across machines, and the system scales efficiently for large-scale ANNS.

\textbf{Load Balancing Strategies.}
Although the initial distribution ensures balanced workloads across machines, pruning introduces imbalance over time.
This occurs because dimensions processed later in the pipeline are more likely to be pruned, as discussed in Section~\ref{subsec:motivation}. 
Resulting in reduced computation for some machines while others experience heavier loads.
To address this, \sysname{} dynamically adjusts the execution order of dimensions to reduce load imbalance.
For example, if the machine responsible for \(D1\) (e.g., \(M1\)) becomes overloaded, subsequent queries (e.g., \(Q2\)) are reconfigured to process \(D1\) last. As pruning is more effective in later stages, this adjustment significantly reduces the burden on overloaded machines. 

In the example shown in Figure~\ref{fig:harmonypipline}(b), \(Q2\)’s dimension \(D1\) is deferred to the final stage, allowing earlier stages to prune unpromising queries and alleviate the load on \(M1\). This adaptive scheduling ensures no machine becomes a bottleneck, maintaining balanced query execution and improving throughput.
\section{Implementation}
\label{implementation}
\begin{figure*}[!h]
  \begin{center}
    \includegraphics[width=1.0\linewidth, height=0.28\linewidth]{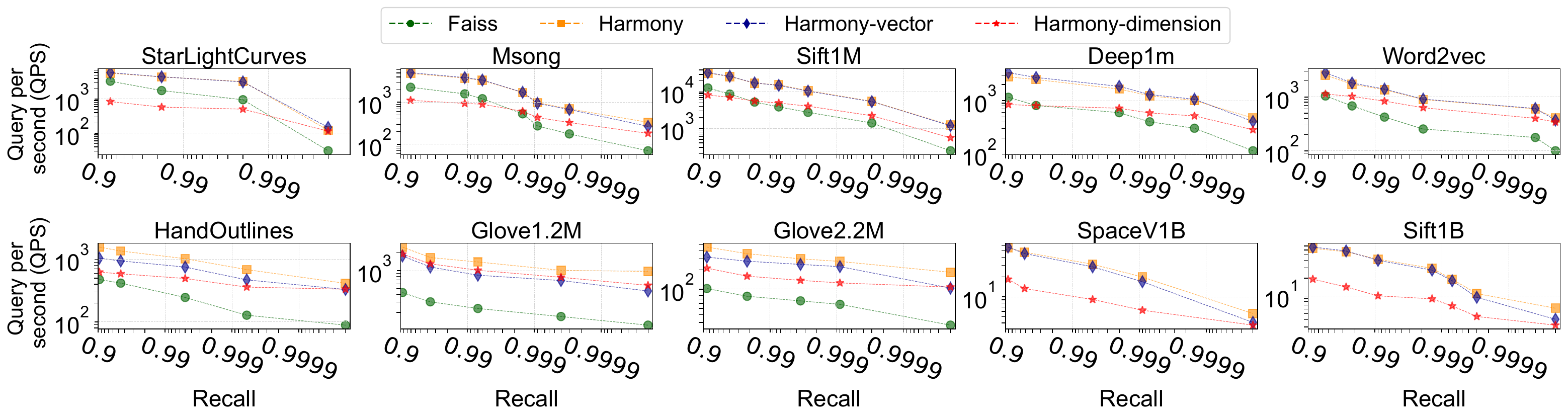}
  \end{center}
  
\vspace{-0.1in}
\vspace{-0.1in}

\caption{Time-accuracy trade-off. }
  \label{fig:overall}
\vspace{-.15in}

\end{figure*}

~\sysname{} is implemented in C++20 with 6000 lines of code. It leverages modern C++ features, such as compile-time computation and zero-cost abstractions. It also employs OpenMPI for inter-node communication and Intel MKL for efficient distance computations.
During query execution, each worker node receives updated top-k heaps as pruning thresholds from the master node.
Once the query parameters are set, workers issue non-blocking (\texttt{MPI\_Isend} / \texttt{MPI\_Irecv}) operations for data and partial results, enabling communication and local computation to overlap. 
Distance calculations within a worker node are parallelized using OpenMP, while Intel MKL accelerates fundamental operations such as L2 or inner-product computations.

\vspace{0.05in}
\noindent
\textbf{Parameters.} To enhance user experience and facilitate ease of use, \sysname{} offers a variety of unique parameters. The parameters are defined as follows:

\begin{itemize}
    \item \texttt{--NMachine} \textit{[nargs=1]}: Specify the number of nodes involved in the distributed index. 
    \item \texttt{--Pruning\_Configuration} \textit{[nargs=0..1]}: Enables or disables vector and dimension-level pruning.
    \item \texttt{--Indexing\_Parameters} \textit{[e.g., nlist, nprobe, dim]}: Control clustering granularity and search scope to balance recall, latency, and memory usage.
    \item \texttt{--$\alpha$} : User-defined parameters to control load balancing and throughput preferences in cost model.
    \item \texttt{--Mode} \textit{[Harmony, Harmony-vector, Harmony-dimension]}: Specify the mode.
\end{itemize}

\section{Evaluation}
\label{sec:evalu}

\subsection{Experiment Setup}

\textbf{Methodology.} We primarily compare \sysname{} against Faiss~\cite{facebook2020a_SOSP_17}, the state-of-the-art stand-alone ANNS query engine, as well as two baseline partitioning methods: pure vector-based partitioning (abbreviated as Harmony-vector) and pure dimension-based partitioning (abbreviated as Harmony-dimension). To ensure a fair comparison, all methods adopt the same clustering algorithm and number of clusters as Faiss. 
After performing the same cluster construction, all the distribution-based strategies distribute the clusters based on their respective distribution strategies (where Harmony is based on the cost model in Section~\ref{subsec:fine-grained}) and perform ANNS. We also compare \sysname{} with Auncel, a distributed ANNS that provides error-bound guarantees. Since high-precision ANNS is the big trend, we mainly focus on the performance difference between Faiss and \sysname{} in high-precision scenarios.

\textbf{Datasets.} 
We use ten open-source datasets of varying sizes and dimensions, as shown in Table~\ref{tab:datasets}. These datasets are created by embedding various types of data (time series, audio, image, word vectors, and text) with different dimensions, providing both data and query vectors. They have been widely evaluated by many ANN systems~\cite{DBLP:conf/osdi/ZhangXCSXCCH00Y23_OSDI_0,Spann} and algorithms~\cite{ADSampling,HNSW,NSG}. We expand the Star Light Curves and Hand Outlines datasets to ensure that they contain sufficient data for distributed processing while satisfying the base distribution.

\begin{table}[h]

\centering
\small
\caption{\textcolor{black}{Dataset statistics.}}

\label{tab:datasets}
\begin{tabular}{@{}lcccc@{}}
\toprule
Dataset & Size & Dim & Query Size & Data Type \\ 
\midrule
Star Light Curves  & 823,600  & 1024 & 1,000 & Time Series \\
Msong   & 992,272  & 420 & 1,000 & Audio \\
Sift1M    & 1,000,000 & 128 & 10,000 & Image \\
Deep1M    & 1,000,000 & 256 & 1,000 & Image \\
Word2vec   & 1,000,000 & 300 & 1,000 & Word Vectors \\
Hand Outlines & 1,000,000 & 2709 & 370 & Time Series \\
Glove1.2m & 1,193,514 & 200 & 1,000 & Text \\
Glove2.2m & 2,196,017 & 300 & 1,000 & Text \\
SpaceV1B   & 1,000,000,000 & 100 & 10,000 & Text \\
Sift1B  & 1,000,000,000 & 128 & 10,000 & Image \\
\bottomrule
\end{tabular}

\end{table}

\textbf{Platform.} We conduct a standard performance evaluation of \sysname{} on a distributed system consisting of 20 nodes. Each node is equipped with an Intel(R) Xeon(R) Gold 6258R CPU, featuring 28 cores and 56 threads, and is connected to 256GB of global memory in a NUMA configuration. The nodes are interconnected via high-speed 100Gb/s networking, and the operating system used is CentOS Stream 8. Each node supports AVX-512, making it suitable for high-performance distributed workloads.

  



\vspace{-.15in}

\subsection{Overall Performance} \label{subsce:overall_performenca}
\subsubsection{QPS-recall trade-off under uniform workloads.}  
In this section, we analyze the trade-off between query per second (QPS) and recall when comparing Faiss with \sysname{}. 
We test \sysname{} on four worker nodes and use Faiss as a baseline on a single node. For two large datasets, SpaceV1B and Sift1B, the single-node solution Faiss and the 4-node solution are unable to handle search requirements due to their size. We employ Harmony with 16 nodes for distributed search. The experimental results are shown in Figure~\ref{fig:overall}.

We have the following three findings.
First, all methods demonstrate scalability, as the three distribution strategies of Harmony outperform Faiss by 3.75$\times$ in terms of average speedup.
Second, under high recall precision conditions, Harmony frequently exceed the theoretical speedup (4 times), achieving a speedup of 4.63$\times$ as pruning reduces the overhead and improves performance.
Third, when the recall is lower than 99\%, the Harmony-Vector partitioning shows the optimal performance. This is because the speedup for Harmony-dimension and Harmony partitioning is limited under lower precision, and the overhead from additional distance computations affects their performance.

\subsubsection{Search performance under skewed workloads.} Due to space constraints, SpaceV1B and Sift1B could not be processed on 4 nodes, and we focus on 8 relatively small datasets in our next experiments.

\begin{figure*}[!h]
  \begin{center}
    \includegraphics[width=1.0\linewidth, height=0.32\linewidth]{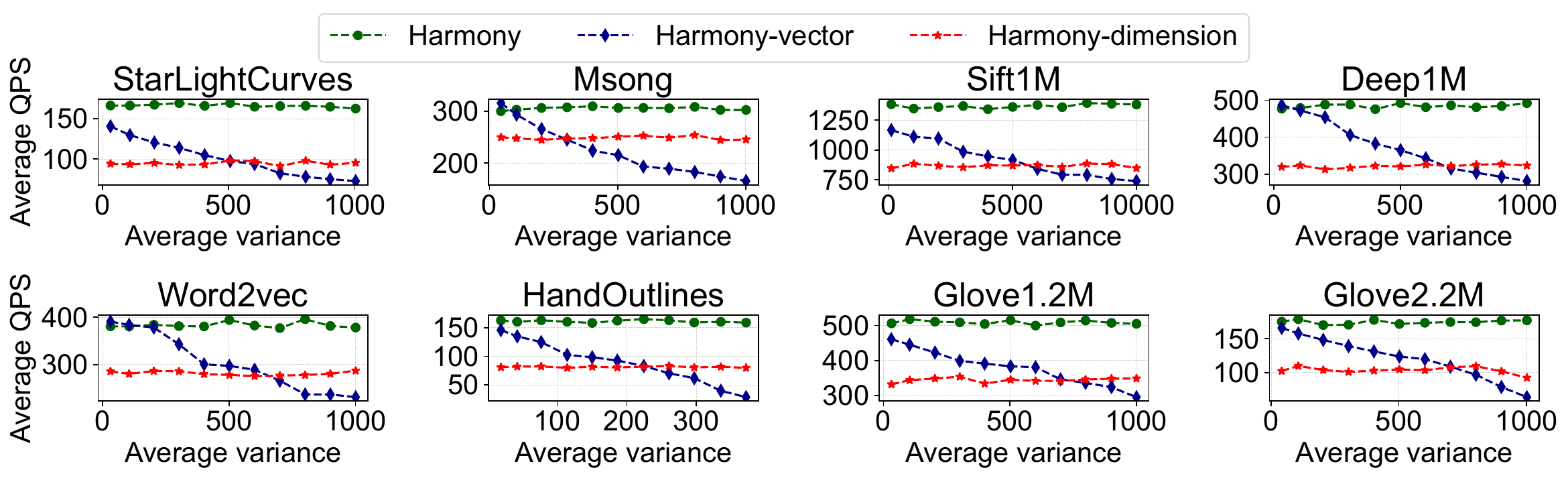}
  \end{center}
  
\vspace{-0.1in}
\vspace{-0.1in}

\caption{Impact of load distribution on query performance. }
  \label{fig:imbalance_load}
\vspace{-.15in}

\end{figure*}

We conduct experiments on four distinct nodes, examining three different strategies of Harmony under varying levels of load imbalance. The query sets are manipulated to ensure different load differences on each machine, and we use the variance (as derived in Section~\ref{subsubsce:cost_mod}) to quantify this imbalance. The final results are shown in Figure~\ref{fig:imbalance_load}.

First, we observe that the traditional vector partition does not adapt well to load imbalance scenarios. As the load becomes more imbalanced, the average QPS decreases 56\% on average.
Second, \sysname{} and \sysname{}-dimension partitioning (dim) both handle imbalance more favorably, showing no clear performance degradation during load skew changes. 
Although the dimension-based approach is more stable due to its absolute partitioning principle, the hybrid approach in Harmony (balancing dim and vector partitioning adaptively) remains advantageous.
Even when the load is extremely imbalanced, Harmony outperforms the dimension-based partitioning by 91\%, demonstrating the advantage of our cost model.
Third, upon further inspection of these datasets, we notice that Harmony retains its advantage consistently for larger-sized vectors or higher feature dimensions. 
As complexity grows and distribution skew worsens, our adaptive strategy in Harmony continues to leverage the synergy between dimension-split and vector-split, thus providing robust performance across varying degrees of imbalance.

\subsection{Performance Ablation Study}  
\subsubsection{Time breakdown.}  
\begin{figure}
    \centering
    \includegraphics[width=\linewidth,height=0.5\linewidth]{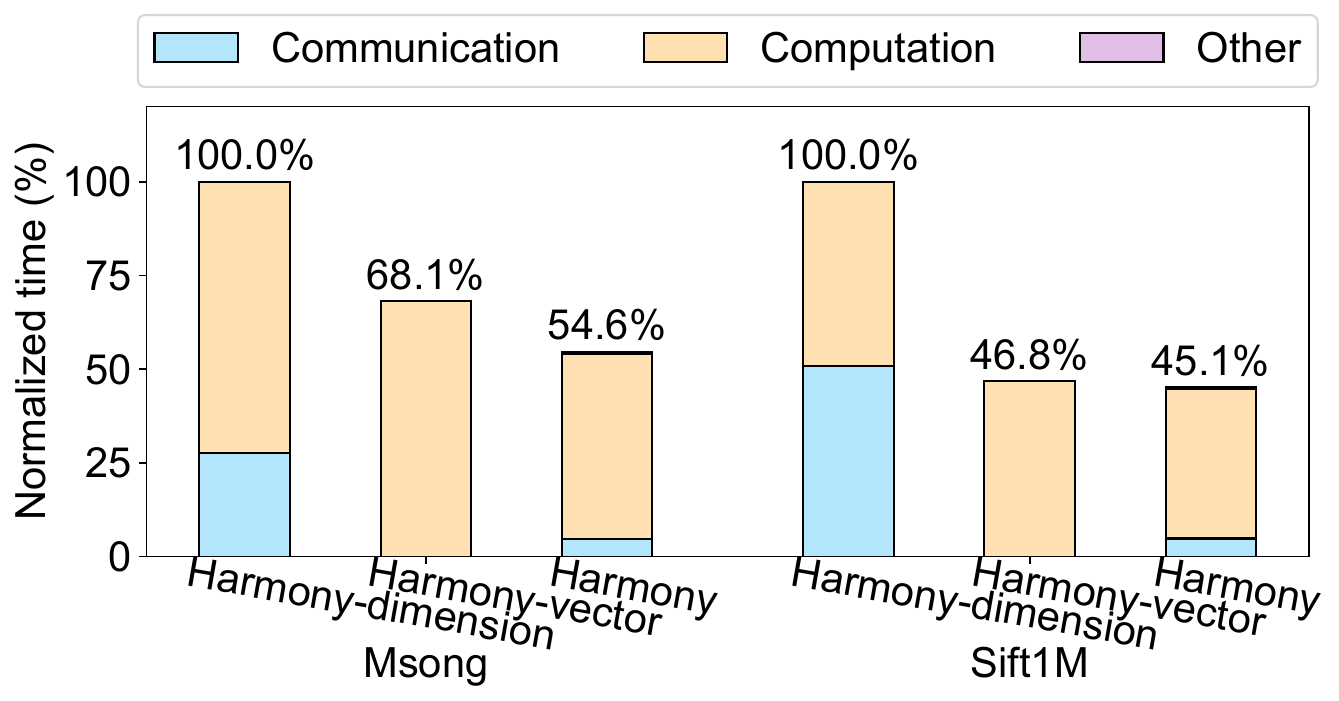}

    \caption{The contribution of the three optimization techniques to \sysname{}'s query throughput.}
    \label{fig:breakdown_evalution}
\vspace{-.25in}
\end{figure}

We analyzed the time spent in various stages of the query process, including data communication, computation, and other overhead. 
The results are shown in Figure~\ref{fig:breakdown_evalution}.

We have the following three key findings. First, except for Harmony-vector, both Harmony and Harmony-dimension incur communication overhead. Specifically, Harmony-dimension has a higher communication overhead due to more dimension slicing. This is evident as the communication time is larger in Harmony-dimension compared to \sysname{}, which has less dimension partitioning. Second, despite the similarity in computation overhead between Harmony-dimension and Harmony-vector, Harmony has a lower computational overhead compared to the other two. This is because Harmony introduces a pruning module, which will be analyzed further in Section~\ref{subsubsec:pruning_break}, where we discuss pruning ratios for each module. Third, the main overheads for all three modules are concentrated in commnuication and computation, and the communication overhead is independent of dimensions, meaning that communication overhead is significantly lower than computation overhead in datasets with larger dimensions. For example, in the 128-dimensional Sift1M dataset, communication overhead is much higher compared to the 420-dimensional Msong dataset.

\subsubsection{Contribution of optimization techniques.}
\begin{figure}
    \centering
    \includegraphics[width=\linewidth,height=0.45\linewidth]{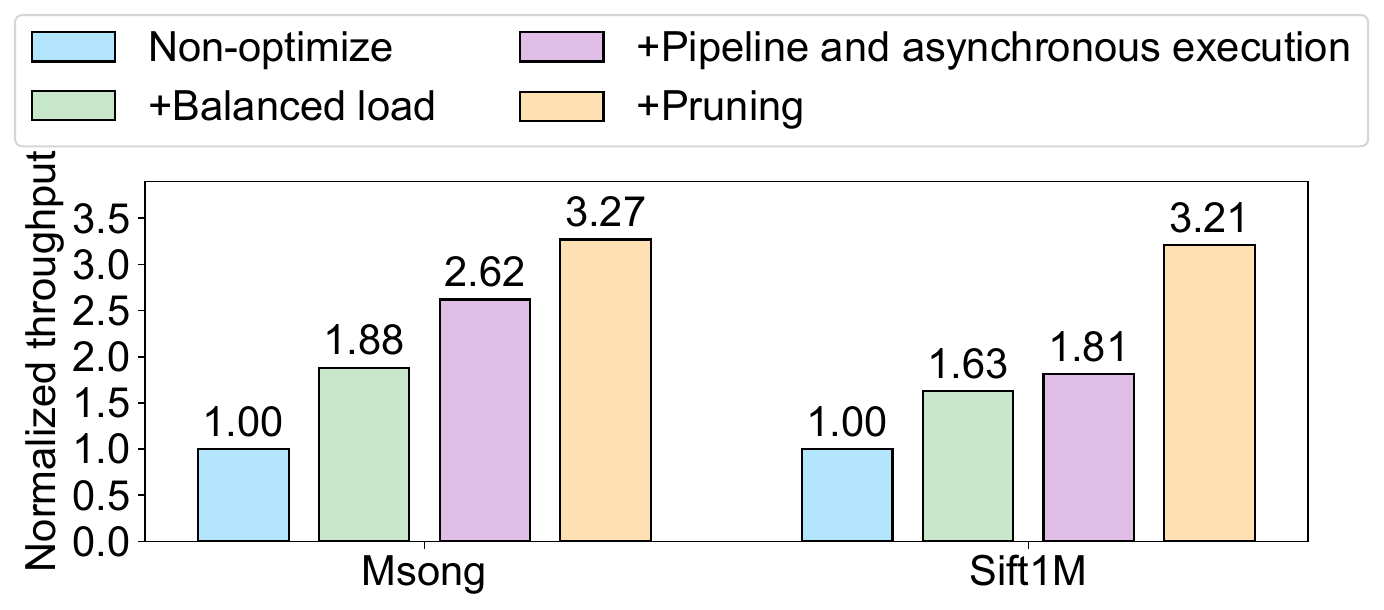}
     \vspace{-0.1in}

    \caption{The contribution of the three optimization techniques to \sysname{}'s query throughput.}
    \label{fig:breakdown_time_query_exp}

\end{figure}

To understand the contribution of each component in \sysname{}, we measure the speedup provided by its key features: balanced load, pipeline and synchronous execution, and pruning in 4 nodes. By isolating these components, we assess their effects on system performance. 
Experimental results are shown in Figure~\ref{fig:breakdown_time_query_exp}. 

We have three main findings. First, on average, each optimization technique provides a performance improvement, with balanced load yielding a 1.75$\times$ increase, pipeline and asynchronous execution achieving a 1.25$\times$ improvement, and pruning offering a 1.51$\times$ enhancement in throughput. 
Second, the performance gain comes from pruning, aligning with the findings in Section~\ref{subsec:motivation}. This demonstrates that pruning is an impactful optimization technique for improving \sysname{}’s query throughput. 
Third, for the Sift1M dataset, the performance improvement from balanced load and pipeline execution is less pronounced. This is due to the relatively uniform distribution of the load, where the benefit of load balancing and asynchronous execution is somewhat diminished. However, pruning still leads to a clear throughput improvement, indicating its robustness in optimizing query performance regardless of data distribution.

\subsubsection{Pruning ratio breakdown.} \label{subsubsec:pruning_break}
We measure the pruning strategy discussed in Section~\ref{subsec:pipelined_engine} to evaluate its effectiveness in reducing computational overhead. In this experiment, we perform a dimensional split of size 4 and apply the pruning strategy to assess its impact on query performance. The results are shown in Table~\ref{tab:Pruning_ratio_comparison}.

\begin{table}[h]
\centering
\small
        \vspace{-0.1in}
\caption{Average pruning ratio comparison across four nodes.}
        \vspace{-0.15in}
\label{tab:Pruning_ratio_comparison}
\begin{tabular}{@{}lcccccc@{}}
\toprule

\textbf{Dataset}  & \textbf{\thead{First \\ Slice (\%)}} &  \textbf{\thead{Second \\ Slice (\%)}} & \textbf{\thead{Third \\ Slice (\%)}} & \textbf{\thead{Fourth \\ Slice (\%)}} & \textbf{\thead{Average \\ Pruning Ratio (\%)}} \\

\midrule
Msong          & 0.00    & 43.14    & 76.06    & 95.29        & 53.87 \\ 
Glove1.2m     & 0.00     & 1.54     & 30.71    & 86.66      & 29.73 \\ 
Word2vec        & 0.00   & 24.85    & 53.77    & 83.66      & 40.32 \\ 
Deep1M         & 0.00    & 7.67     & 66.09    & 97.36     & 42.03 \\ 
Sift1m        & 0.00     & 41.76    & 85.04    & 98.40     & 57.05 \\ 
Star          & 0.00     & 81.24    & 95.23    & 99.05   & 69.14 \\ 
Glove2.2m    & 0.00      & 5.14     & 30.70    & 81.18     & 29.76 \\ 
Hand          & 0.00     & 63.54    & 91.62    & 98.10     & 63.83 \\ 

\bottomrule
\end{tabular}
        \vspace{-0.15in}
\end{table}

We have the following three observations. First, later slices exhibit better pruning rates. Specifically, the average pruning ratio for the second slice is 33.61\%, for the third slice is 66.15\%, and for the fourth slice is 92.33\%. This is because subsequent query slices benefit from the results of earlier slices. As discussed in Section~\ref{subsec:motivation}, later slices can achieve better pruning thresholds due to the accumulation of query results.
Second, pruning rates vary significantly across different datasets. For instance, in the Glove1.2M dataset, the third slice achieves only a 30.71\% pruning rate, while in the Star dataset, it reaches 95.23\%. 
This variance is mainly due to the differences in dataset distributions, where some datasets are more easily pruned than others.
Third, the pruning rate in the final slice consistently exceeds 80\%. 
This indicates that the vast majority of distance computations do not require the use of the last dimension, which aligns with our analysis in Section~\ref{subsec:motivation}. 
This highlights that most vectors are pruned by the time the final slice is reached.

\subsection{Index Build Experiments}  

\subsubsection{Index build time.}
\begin{figure*}
    \centering
    \includegraphics[width=\linewidth,height=0.185\linewidth]{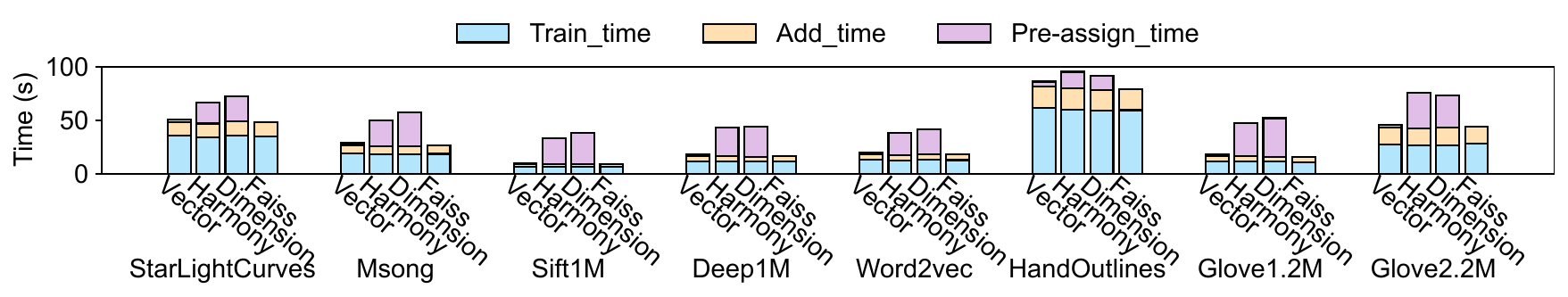}
\vspace{-0.13in}
    \caption{\sysname{}'s index build time breakdown.}
    \label{fig:breakdown_build_time}

\end{figure*}

We measure the time and space overhead of constructing indexes using \sysname{} and Faiss. 
First, we focus on the time overhead. 
We measure three distribution methods: Harmony-vector (abbreviated as Vector), Harmony-dimension (abbreviated as Dimension) indexing in 4 nodes environment, and \sysname{}, along with Faiss indexing in a single-machine environment. 
The experimental results are shown in Figure~\ref{fig:breakdown_build_time}, where we break down the index construction into three stages: training the clustering centers (Train), distributing the base vectors to the clustering centers (Add), and the distributed-specific operation of assigning parts of the index to different machines (Pre-assign).

We have the following three observations. 
First, the train and add times for all methods are similar, as \sysname{} does not require modifying the index structure and exhibits good scalability. 
Second, the pre-assign time for Harmony-dimension and \sysname{} is longer compared to Harmony-vector. This is because these two distribution methods first allocate space and initialize intermediate results related to distance computation, which takes some time and depends on the data size. For example, on the Glove2.2m dataset, the time is approximately twice as long as that of Glove1.2m. 
Third, for datasets of the same size, the train and add times are generally proportional to the dimensionality of the dataset. This is because both training and distribution scale linearly with dimensionality.

\subsubsection{Index space overhead.}  \label{subsubsec:index_memory}
In this experiment, we compare the space overhead of \sysname{}'s index with Faiss and baseline configurations. Efficient use of memory is critical for large-scale ANNS systems, and this experiment quantifies the storage efficiency of \sysname{}'s hybrid partitioning and pruning mechanisms.  
The results are shown in Table~\ref{tab:index_sizes}.

\begin{table}[h]
\centering
\small
\caption{\textcolor{black}{Index memory comparison.}}
\label{tab:index_sizes}
\begin{tabular}{@{}lcccc@{}}
\toprule
\textbf{Dataset}            & \textbf{Faiss}    & \textbf{\thead{Harmony \\ vector}}   & \textbf{\thead{Harmony \\ dimension}}   & \textbf{Harmony}     \\ 
\midrule
StarLightCurves     & 3.2GB            & 788MB             & 815MB                & 798MB        \\
Msong               & 1.6GB            & 411MB             & 418MB                & 413MB        \\
Sift1M              & 497MB            & 126MB             & 131MB                & 128MB        \\
Deep1M              & 986MB            & 245MB             & 253MB                & 250MB        \\
Word2Vec            & 1.2GB            & 258MB             & 295MB                & 279MB        \\
HandOutlines        & 6.1GB            & 1.50GB            & 1.54GB               & 1.51GB       \\
Glove1.2M           & 921MB            & 227MB             & 238MB                & 233MB        \\
Glove2.2M           & 2.5GB            & 660MB             & 697MB                & 686MB        \\
\bottomrule
\end{tabular}
\end{table}

We identify the following three key points. First, the index overhead is proportional to both the dataset size and its dimensions. For example, the space used by Msong is 3.2 times larger than that of Sift1M, which aligns with their respective dataset sizes (dimension $\times$ dataset size). Second, all three partitioning schemes occupy about 1/4 of the space of Faiss (single-machine index), suggesting that distributed indexing does not require significant additional space overhead. Third, both Harmony and Harmony-dimension incur some additional space overhead, but this overhead is minimal, representing only about 2\% of the original space size.

\vspace{-0.1in}
\subsection{Additional Discussion}

\begin{figure}[t]
    \centering
    \begin{subfigure}[t]{0.52\linewidth}
        \includegraphics[width=\linewidth]{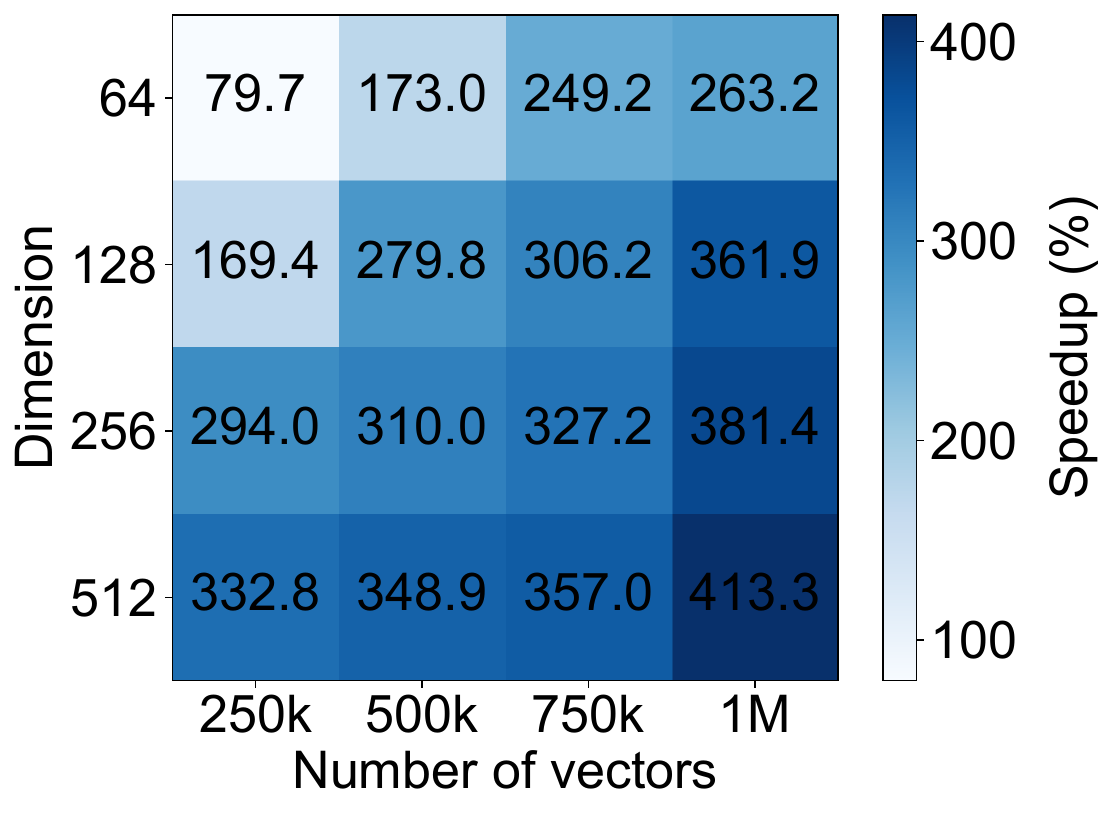}
        \caption{Relationship between pruning ratio and search probes}
        \label{fig:sub_distance}
    \end{subfigure}
    \hfill
    \begin{subfigure}[t]{0.47\linewidth}
        \includegraphics[width=\linewidth]{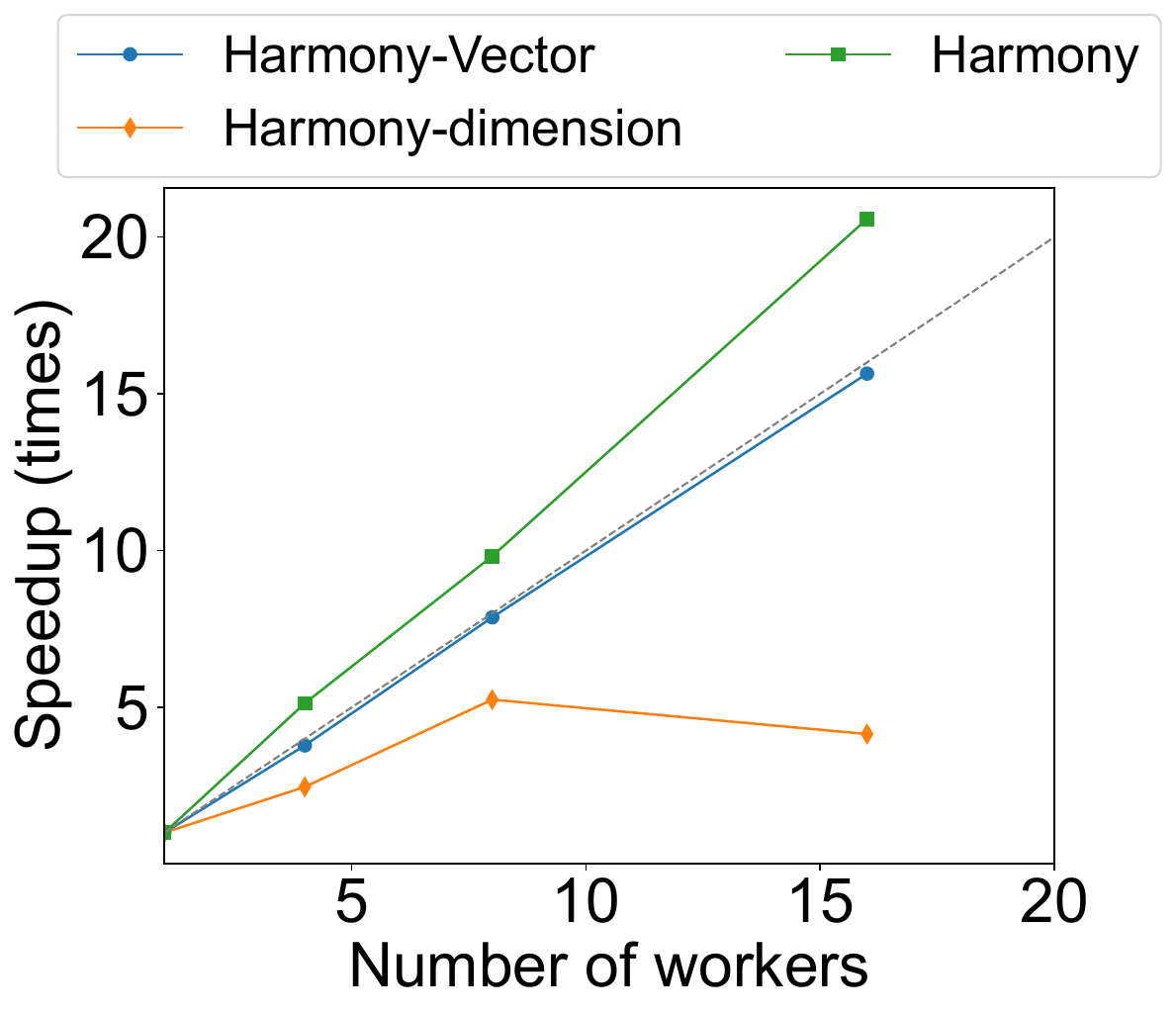}
        \caption{Results of subNN indexes with different parameters.}
        \label{fig:sub_angle}
    \end{subfigure}
    \caption{Impact of pruning effectiveness and indexing parameters on search performance}
    \label{fig:Discussion_Construct}    
    \vspace{-0.1in}

\end{figure}

\subsubsection{Impact of dimensions and dataset size on harmony performance.}
We construct datasets that follow a Gaussian distribution, with dimensions ranging from 64 to 512 and sizes varying from 250K to 1M, using four nodes. The speedup achieved by Harmony is shown in Figure~\ref{fig:Discussion_Construct} (a).

We have three key findings. First, as both the dataset size and dimension increase, performance improves. Specifically, for each doubling of the dimension, the speedup increases by 26.8\%, and for each doubling of the dataset size, the speedup increases by 25.9\%. Second, in large datasets (1M, 512 dimensions), Harmony shows a speedup greater than the number of machines. This is due to the introduction of pruning techniques, which reduce computational overhead. Third, Harmony’s performance is suboptimal for smaller datasets, which is attributed to the increased communication overhead that dominates the overall performance in such cases.

\subsubsection{Scalability.}  
To validate the scalability of \sysname{}, we perform experiments testing its performance across different machine configurations: 4, 8, and 16 nodes. The speedup achieved by Harmony is shown in Figure~\ref{fig:Discussion_Construct} (b).

We have three key findings. First, the group-based partitioning method consistently demonstrates a speedup greater than the number of machines, owing to the pruning design, which effectively reduces computational overhead. 
Second, the vector-based partitioning achieves performance close to the number of workers, as it does not introduce additional overhead and scales linearly with the number of workers. 
Third, the dimension-based partitioning method exhibits an increase in performance as the number of workers grows, followed by a decline. This occurs because as the number of dimensions increases, communication overhead grows, which eventually reduces the overall performance gain.

\subsubsection{Peak memory usage}  
To evaluate memory efficiency during query execution, we measure the peak memory consumption of \sysname{} and compare it to baseline partition methods in 4 nodes.  
The results are shown in Table~\ref{tab:Peak_mem}. 

\begin{table}[h]
\centering
\small
\caption{\textcolor{black}{Peak memory comparison.}}
    \vspace{-0.1in}
\label{tab:Peak_mem}
\begin{tabular}{@{}lccc@{}}
\toprule
Dataset           & \sysname{}-vector   & \sysname{}   & \sysname{}-dimension  \\ 
\midrule
StarLightCurves   & 3.94GB   & 4.01GB    & 4.07GB    \\
Msong             & 1.15GB   & 1.32GB    & 1.46GB    \\
Sift1M            & 1.37GB   & 1.72GB   & 1.96GB   \\
Deep1M            & 1.23GB   & 1.61GB   & 1.88GB    \\
Word2Vec          & 658MB     & 723MB    & 812MB    \\
HandOutlines      & 11.06GB  & 11.19GB    & 11.33GB    \\
Glove1.2M         & 814MB     & 932MB    & 1.06GB    \\
Glove2.2M         & 1.64GB   & 1.98GB    & 2.23GB   \\
\bottomrule
\end{tabular}
     \vspace{-0.1in}
\end{table}

We observe the following three key findings. First, the peak memory usage for the three partition methods is generally proportional to both the data dimensions and dataset size. Second, the peak memory consumption for \sysname{} and \sysname{}-dimension is slightly higher than that of \sysname{}-vector, but this increase diminishes as the data dimension grows. For instance, in the case of Deep1M with 100 dimensions, \sysname{} exceeds \sysname{}-vector by 30.9\%, whereas in HandOutlines with 2709 dimensions, the increase drops to 1.17\%. This behavior occurs because all methods involving dimension partitioning require additional intermediate results unrelated to the data dimensions, and these results become diluted as the dimensions increase, which demonstrates the potential of \sysname{} in handling high-dimensional data. Third, the increase in memory usage due to intermediate results is related to the number of dimension slices. For example, \sysname{} shows a smaller increase in memory consumption compared to \sysname{}-dimension because it does not partition entirely based on dimensions.

\subsubsection{Comparison with Auncel and \sysname{}.}
Auncel~\cite{NSDI2023} focuses on handling distributed vector queries with low search load under specific precision requirements. It uses a fixed partitioning strategy similar to Harmony-vector, distributing the query load across the system. However, Auncel's performance in load balancing is not optimal under skewed workloads, which is similar to the behavior of Harmony-vector. 

In contrast, Harmony addresses skewed workloads by leveraging pruning techniques and fine-grained load balancing, allowing it to better handle uneven query distributions. While both systems share a similar partitioning approach, Harmony outperforms Auncel in scenarios involving load imbalance, as it can take advantage of pruning opportunities to maintain throughput and stability.




\flushbottom
\onecolumn
\begin{multicols}{2}
\raggedcolumns

\bibliographystyle{plain}

\bibliography{reference}

\begin{thebibliography}{10}

\bibitem{ApacheDrill}
Apache drill.
\newblock \url{https://drill.apache.org/}.
\newblock Accessed: 2025-01-08.

\bibitem{CockroachDB}
Cockroachdb.
\newblock \url{https://www.cockroachlabs.com/product/cockroachdb/}.
\newblock Accessed: 2025-01-08.

\bibitem{PrestoSparkSQL}
Integrating presto with spark sql.
\newblock \url{https://prestodb.io/docs/current/connector/spark.html}.
\newblock Accessed: 2025-01-08.

\bibitem{Neo4j}
Neo4j.
\newblock \url{https://neo4j.com/}.
\newblock Accessed: 2025-01-08.

\bibitem{distribute2}
Pinecone.
\newblock \url{http://pinecone.io}.
\newblock Accessed: 2025-01-08.

\bibitem{Presto}
Presto: Distributed sql query engine for big data.
\newblock \url{https://prestodb.io/}.
\newblock Accessed: 2025-01-08.

\bibitem{distribute4}
Qdrant.
\newblock \url{http://qdrant.tech}.
\newblock Accessed: 2025-01-08.

\bibitem{TigerGraph}
Tigergraph.
\newblock \url{https://www.tigergraph.com/}.
\newblock Accessed: 2025-01-08.

\bibitem{distribute1}
Vald.
\newblock \url{http://vald.vdaas.org}.
\newblock Accessed: 2025-01-08.

\bibitem{distribute3}
Vespa.
\newblock \url{https://vespa.ai/}.
\newblock Accessed: 2025-01-08.

\bibitem{JanusGraph}
Janusgraph.
\newblock \url{https://janusgraph.org/}, 2017.
\newblock Accessed: 2025-01-08.

\bibitem{UCI}
{UCI} machine learning repository.
\newblock \url{https://archive.ics.uci.edu/}, 2024.

\bibitem{youtube}
Youtube.
\newblock \url{https://www.youtube.com/}, 2024.

\bibitem{babenko2014a_OSDI_17}
Artem Babenko, Victor Lempitsky, B.Hari Babu, N.Subhash Chandra, and T.V. Gopal.
\newblock The inverted multi-index.
\newblock {\em IEEE transactions on pattern analysis and machine intelligence}, 37(6):1247–1260, 2014.

\bibitem{baranchuk2018a_OSDI_19}
Dmitry Baranchuk, Artem Babenko, and Yury Malkov.
\newblock Revisiting the inverted indices for billion-scale approximate nearest neighbors.
\newblock In {\em Proceedings ofthe European Conference on Computer Vision (ECCV}, pages 202–216,, 2018.

\bibitem{bentley1975a_OSDI_23}
Jon~Louis Bentley.
\newblock Multidimensional binary search trees used for associative searching.
\newblock {\em Communications ofthe ACM}, 18(9):509–517, 1975.

\bibitem{Spann}
Qi~Chen, Bing Zhao, Haidong Wang, Mingqin Li, Chuanjie Liu, Zengzhong Li, Mao Yang, and Jingdong Wang.
\newblock Spann: Highly-efficient billion-scale approximate nearest neighborhood search.
\newblock In M.~Ranzato, A.~Beygelzimer, Y.~Dauphin, P.S. Liang, and J.Wortman Vaughan, editors, {\em Advances in Neural Information Processing Systems}, volume~34, page 5199–5212. Curran Associates, Inc, 2021.

\bibitem{clarkson-a_OSDI_28}
Kenneth~L. Clarkson.
\newblock An algorithm for approximate closest-point queries.
\newblock In {\em Proceedings ofthe tenth annual symposium on Computational geometry}, pages 160–164,1994.

\bibitem{Spanner}
James~C. Corbett, Jeffrey Dean, Michael Epstein, Andrew Fikes, Christopher Frost, J.~J. Furman, Sanjay Ghemawat, Andrey Gubarev, Christian Heiser, Peter Hochschild, and et~al.
\newblock {Spanner}: {Google}'s globally-distributed database.
\newblock In {\em Proceedings of the 10th USENIX Symposium on Operating Systems Design and Implementation (OSDI)}, pages 251--264, 2012.

\bibitem{datar2004a_OSDI_30}
Mayur Datar, Nicole Immorlica, Piotr Indyk, and Va-hab~S. Mirrokni.
\newblock Locality-sensitive hashing scheme based on p-stable distributions.
\newblock In {\em Proceedings of the Twentieth Annual Symposium on Computational Geometry, SCG ’04}, pages 253–262,, 2004.

\bibitem{dean2013a_OSDI_31}
Jeffrey Dean and Luiz~André Barroso.
\newblock The tail at scale.
\newblock {\em Communications of the ACM}, 56(2):74–80, 2013.

\bibitem{dean2008mapreduce}
Jeffrey Dean and Sanjay Ghemawat.
\newblock Mapreduce: Simplified data processing on large clusters.
\newblock In {\em Communications of the ACM}, volume~51, pages 107--113. ACM New York, NY, USA, 2008.

\bibitem{delaunay1934a_SOSP_15}
B.N. Delaunay.
\newblock Sur la sphère vide.
\newblock {\em Bull. Acad. Sci. URSS}, 6:793–800, 1934.

\bibitem{dong2011a_SOSP_16}
Wei Dong, Moses Charikar, and Kai Li.
\newblock Efficient k-nearest neighbor graph construction for generic similarity measures.
\newblock In {\em Proceedings of the 20th International Conference on World Wide Web, WWW2011}, page 577–586, Hyderabad, India, 2011. Association for Computing Machinery.

\bibitem{facebook2020a_SOSP_17}
Facebook.
\newblock Faiss, 2020.
\newblock Accessed: 2024-10-15.

\bibitem{NSG}
Cong Fu, Chao Xiang, Changxu Wang, and Deng Cai.
\newblock Fast approximate nearest neighbor search with the navigating spreading-out graph.
\newblock {\em Proc. VLDB Endow.}, 12(5):461–474, jan 2019.

\bibitem{gabriel1969a_SOSP_20}
K.Ruben Gabriel and Robert~R. Sokal.
\newblock A new statistical approach to geographic variation analysis.
\newblock {\em Systematic zoology}, 18, 3:259–278, 1969.

\bibitem{ADSampling}
Jianyang Gao and Cheng Long.
\newblock High-dimensional approximate nearest neighbor search: with reliable and efficient distance comparison operations.
\newblock {\em Proc. ACM Manag. Data}, 1(2), jun 2023.

\bibitem{RaBitQ}
Jianyang Gao and Cheng Long.
\newblock Rabitq: Quantizing high-dimensional vectors with a theoretical error bound for approximate nearest neighbor search.
\newblock {\em Proc. ACM Manag. Data}, 2(3), may 2024.

\bibitem{gray2000a_OSDI_36}
Wayne~D. Gray and Deborah~A. Boehm-Davis.
\newblock Milliseconds matter: An introduction to microstrategies and to their use in describing and predicting interactive behavior.
\newblock {\em Journal of experimental psychology: applied}, 6(4), 2000.

\bibitem{guo2022manu}
Rentong Guo, Xiaofan Luan, Long Xiang, Xiao Yan, Xiaomeng Yi, Jigao Luo, Qianya Cheng, Weizhi Xu, Jiarui Luo, Frank Liu, et~al.
\newblock Manu: a cloud native vector database management system.
\newblock {\em arXiv preprint arXiv:2206.13843}, 2022.

\bibitem{hajebi2011a_SOSP_23}
Kiana Hajebi, Yasin Abbasi-Yadkori, Hossein Shahbazi, and Hong Zhang.
\newblock Fast approximate nearest-neighbor search with k-nearest neighbor graph.
\newblock In {\em IJCAI 2011, Proceedings ofthe 22nd International Joint Conference on Artificial Intelligence}, Barcelona, Catalonia, Spain, 2011.
\newblock AAAI Press, 1312–1317.

\bibitem{jain2008a_OSDI_41}
P.~Jain, B.~Kulis, and K.~Grauman.
\newblock Fast image search for learned metrics.
\newblock In {\em 2008 IEEE Conference on Computer Vision and Pattern Recognition}, pages 1–8,, 2008-06.

\bibitem{jalaparti2013a_OSDI_42}
Virajith Jalaparti, Peter Bodik, Srikanth Kandula, Ishai Menache, Mikhail Rybalkin, and Chenyu Yan.
\newblock Speeding up distributed request-response workflows.
\newblock {\em ACM SIG-COMM Computer Communication Review}, 43(4):219–230, 2013.

\bibitem{diskann}
Suhas Jayaram~Subramanya, Fnu Devvrit, Harsha~Vardhan Simhadri, Ravishankar Krishnawamy, and Rohan Kadekodi.
\newblock Diskann: Fast accurate billion-point nearest neighbor search on a single node.
\newblock In H.~Wallach, H.~Larochelle, A.~Beygelzimer, F.~d\textquotesingle Alch\'{e}-Buc, E.~Fox, and R.~Garnett, editors, {\em Advances in Neural Information Processing Systems}, volume~32. Curran Associates, Inc., 2019.

\bibitem{jegou2010a_OSDI_44}
Herve Jegou, Matthijs Douze, and Cordelia Schmid.
\newblock Product quantization for nearest neighbor search.
\newblock {\em IEEE transactions on pattern analysis and machine intelligence}, 33(1):117–128, 2010.

\bibitem{j2011a_OSDI_45}
Hervé Jégou, Romain Tavenard, Matthijs Douze, and Laurent Amsaleg.
\newblock Searching in one billion vectors: rerank with source coding.
\newblock In {\em 2011 IEEE International Conference on Acoustics, Speech and Signal Processing (ICASSP}, page 861–864. IEEE, 2011.

\bibitem{kalantidis2014a_OSDI_48}
Yannis Kalantidis and Yannis Avrithis.
\newblock Locally optimized product quantization for approximate nearest neighbor search.
\newblock In {\em Proceedings of the IEEE Conference on Computer Vision and Pattern Recognition (CVPR}, pages 2321–2328,, 2014.

\bibitem{koenigstein2012a_OSDI_49}
Noam Koenigstein, Parikshit Ram, and Yuval Shavitt.
\newblock Efficient retrieval of recommendations in a matrix factorization framework.
\newblock In {\em Proceedings of the 21st ACM international conference on Information and knowledge management}, pages 535–544,, 2012.

\bibitem{kulis2009a_SOSP_32}
Brian Kulis and Kristen Grauman.
\newblock Kernelized locality-sensitive hashing for scalable image search.
\newblock In {\em Computer Vision, 2009 IEEE 12th International Conference on}, page 2130–2137, V., NLD, 2009. IEEE, Elsevier Science Publishers B.

\bibitem{li2017a_OSDI_53}
Hui Li, Tsz~Nam Chan, Man~Lung Yiu, and Nikos Mamoulis.
\newblock Fexipro: fast and exact inner product retrieval in recommender systems.
\newblock In {\em Proceedings of the 2017 ACM International Conference on Management of Data}, pages 835–850,, 2017.

\bibitem{li2018a_SOSP_34}
Jie Li, Haifeng Liu, Chuanghua Gui, Jianyu Chen, Zhenyuan Ni, Ning Wang, and Yuan Chen.
\newblock The design and implementation ofa real time visual search system on jd e-commerce platform.
\newblock In {\em Proceedings ofthe 19th International Middleware Conference Industry (Rennes, France}, page 9–16, New York, NY, USA, 2018. Association for Computing Machinery.

\bibitem{li2021a_SOSP_35}
Sen Li, Fuyu Lv, Taiwei Jin, Guli Lin, Keping Yang, Xiaoyi Zeng, XiaoMing Wu, and Qianli Ma.
\newblock Embedding-based product retrieval in taobao search.
\newblock In {\em Proceedings ofthe 27th ACMSIGKDD Conference on Knowledge Discovery \& Data Mining (KDD ’21}, page 3181–3189, New York, NY, USA, 2021. Association for Computing Machinery.

\bibitem{lian2020a_OSDI_56}
Defu Lian, Haoyu Wang, Zheng Liu, Jianxun Lian, Enhong Chen, and Xing Xie.
\newblock Lightrec: A memory and search-efficient recommender system.
\newblock In {\em Proceedings of The Web Conference 2020}, pages 695–705,, 2020.

\bibitem{liu2004a_OSDI_57}
Ting Liu, Andrew~W. Moore, Alexander Gray, and Ke~Yang.
\newblock An investigation of practical approximate nearest neighbor algorithms.
\newblock In {\em Advances in Neural Information Processing Systems 17 [Neural Information Processing Systems, {NIPS} 2004}, pages 825–832,. Vancouver, British Columbia, Canada, December 13-18, 2004.

\bibitem{Distribute_VB}
Fangzhou~Alec Lu.
\newblock Towards software-defined fpga acceleration for big data analytics.
\newblock 2024.

\bibitem{HNSW}
Yu~A. Malkov and D.~A. Yashunin.
\newblock Efficient and robust approximate nearest neighbor search using hierarchical navigable small world graphs.
\newblock {\em IEEE Trans. Pattern Anal. Mach. Intell.}, 42(4):824–836, apr 2020.

\bibitem{mikolov2013a_SOSP_42}
Tomas Mikolov, Kai Chen, Greg Corrado, and Jeffrey Dean.
\newblock Efficient estimation of word representations in vector space, 2013.
\newblock arXiv:1301.3781 [cs.CL.

\bibitem{muja2014a_OSDI_62}
Marius Muja and David~G. Lowe.
\newblock Scalable nearest neighbour algorithms for high dimensional data.
\newblock {\em IEEE Transactions on Pattern Analysis and Machine Intelligence}, 36(11):2227–2240, 2014.

\bibitem{nishtala2013a_OSDI_64}
Rajesh Nishtala, Hans Fugal, Steven Grimm, Marc Kwiatkowski, Herman Lee, Harry~C. Li, Ryan McElroy, Mike Paleczny, Daniel Peek, and Paul Saab.
\newblock Scaling memcache at facebook.
\newblock In {\em 10th USENIX Symposium on Networked Systems Design and Implementation (NSDI 13}, pages 385–398,, 2013.

\bibitem{DBLP:conf/ppopp23}
Zhen Peng, Minjia Zhang, Kai Li, Ruoming Jin, and Bin Ren.
\newblock iqan: Fast and accurate vector search with efficient intra-query parallelism on multi-core architectures.
\newblock In Maryam~Mehri Dehnavi, Milind Kulkarni, and Sriram Krishnamoorthy, editors, {\em Proceedings of the 28th {ACM} {SIGPLAN} Annual Symposium on Principles and Practice of Parallel Programming, PPoPP 2023, Montreal, QC, Canada, 25 February 2023 - 1 March 2023}, pages 313--328. {ACM}, 2023.

\bibitem{pennington2014a_SOSP_49}
Jeffrey Pennington, Richard Socher, and Christopher Manning.
\newblock Glove: Global vectors for word representation.
\newblock {\em EMNLP}, 14:1532–1543, 2014.

\bibitem{ren2020a_SOSP_51}
Jie Ren, Minjia Zhang, and Dong Li.
\newblock Hm-ann: Efficient billionpoint nearest neighbor search on heterogeneous memory.
\newblock {\em Proceedings ofthe 34th International Conference on Neural Information Processing Systems}, 895:20, 2020.

\bibitem{F1}
Jeff Shute and et~al.
\newblock F1: A distributed sql database that scales.
\newblock In {\em Proceedings of the VLDB Endowment}, volume~6, pages 1068--1079, 2013.

\bibitem{toussaint1980a_SOSP_60}
Godfried~T. Toussaint.
\newblock The relative neighbourhood graph of a finite planar set.
\newblock {\em Pattern recognition}, 12, 4:261–268, 1980.

\bibitem{wang2021a_SOSP_65}
Jianguo Wang, Xiaomeng Yi, Rentong Guo, Hai Jin, Peng Xu, Shengjun Li, Xiangyu Wang, Xiangzhou Guo, Chengming Li, Xiaohai Xu, Kun Yu, Yuxing Yuan, Yinghao Zou, Jiquan Long, Yudong Cai, Zhenxiang Li, Zhifeng Zhang, Yihua Mo, Jun Gu, Ruiyi Jiang, Yi~Wei, and Charles Xie.
\newblock Milvus: A purpose-built vector data management system.
\newblock In {\em Proceedings ofthe 2021 International Conference on Management ofData (Virtual Event, China) (SIGMOD ’21}, page 2614–2627, New York, NY, USA, 2021. Association for Computing Machinery.

\bibitem{wang2012a_SOSP_63}
Jing Wang, Jingdong Wang, Gang Zeng, Zhuowen Tu, Rui Gan, and Shipeng Li.
\newblock Scalable k-nn graph construction for visual descriptors.
\newblock In {\em Computer Vision and Pattern Recognition (CVPR), 2012 IEEE Conference on}, USA, 2012. IEEE, IEEE Computer Society.

\bibitem{wang2014a_OSDI_78}
Jingdong Wang, Naiyan Wang, You Jia, Jian Li, Gang Zeng, Hongbin Zha, and Xian~Sheng Hua.
\newblock Trinary-projection trees for approximate nearest neighbor search.
\newblock {\em IEEE Transactions on Pattern Analysis and Machine Intelligence}, 36(2):388–403, 2014.

\bibitem{wang2018a_OSDI_79}
Jingdong Wang, Ting Zhang, Jingkuan Song, Nicu Sebe, and Heng~Tao Shen.
\newblock A survey on learning to hash.
\newblock {\em IEEE Transactions on Pattern Analysis and Machine Intelligence}, 40(4):769–790, 2018.

\bibitem{weiss2009a_OSDI_81}
Yair Weiss, Antonio Torralba, and Rob Fergus.
\newblock Spectral hashing.
\newblock In {\em Advances in neural information processing systems}, pages 1753–1760,. 2009.

\bibitem{white2012hadoop}
Tom White.
\newblock {\em Hadoop: The Definitive Guide}.
\newblock O'Reilly Media, Inc., 3rd edition, 2012.

\bibitem{xiao2022a_OSDI_84}
Shitao Xiao, Zheng Liu, Weihao Han, Jianjin Zhang, Yingxia Shao, Defu Lian, Chaozhuo Li, Hao Sun, Denvy Deng, and Liangjie Zhang.
\newblock Progressively optimized bi-granular document representation for scalable embedding based retrieval.
\newblock In {\em Proceedings ofthe ACMWeb Conference 2022}, pages 286–296,, 2022.

\bibitem{DBLP:conf/sosp/XuLLXCZLYYYCY23_SOSP_0}
Yuming Xu, Hengyu Liang, Jin Li, Shuotao Xu, Qi~Chen, Qianxi Zhang, Cheng Li, Ziyue Yang, Fan Yang, Yuqing Yang, Peng Cheng, and Mao Yang.
\newblock Spfresh: Incremental in-place update for billion-scale vector search.
\newblock In Jason Flinn, Margo~I. Seltzer, Peter Druschel, Antoine Kaufmann, and Jonathan Mace, editors, {\em Proceedings of the 29th Symposium on Operating Systems Principles, {SOSP} 2023, Koblenz, Germany, October 23-26, 2023}, pages 545--561. {ACM}, 2023.

\bibitem{zaharia2010spark}
Matei Zaharia, Mosharaf Chowdhury, Michael~J Franklin, Scott Shenker, and Ion Stoica.
\newblock Spark: Cluster computing with working sets.
\newblock In {\em Proceedings of the 2nd USENIX conference on Hot topics in cloud computing (HotCloud)}, volume~10, pages 10--10, 2010.

\bibitem{FPGA}
Shulin Zeng, Zhenhua Zhu, Jun Liu, Haoyu Zhang, Guohao Dai, Zixuan Zhou, Shuangchen Li, Xuefei Ning, Yuan Xie, Huazhong Yang, and Yu~Wang.
\newblock Df-gas: a distributed fpga-as-a-service architecture towards billion-scale graph-based approximate nearest neighbor search.
\newblock In {\em Proceedings of the 56th Annual IEEE/ACM International Symposium on Microarchitecture}, MICRO '23, page 283–296, New York, NY, USA, 2023. Association for Computing Machinery.

\bibitem{DBLP:conf/osdi/ZhangXCSXCCH00Y23_OSDI_0}
Qianxi Zhang, Shuotao Xu, Qi~Chen, Guoxin Sui, Jiadong Xie, Zhizhen Cai, Yaoqi Chen, Yinxuan He, Yuqing Yang, Fan Yang, Mao Yang, and Lidong Zhou.
\newblock {VBASE:} unifying online vector similarity search and relational queries via relaxed monotonicity.
\newblock In Roxana Geambasu and Ed~Nightingale, editors, {\em 17th {USENIX} Symposium on Operating Systems Design and Implementation, {OSDI} 2023, Boston, MA, USA, July 10-12, 2023}, pages 377--395. {USENIX} Association, 2023.

\bibitem{zhang2014a_OSDI_90}
Ting Zhang, Chao Du, and Jingdong Wang.
\newblock Composite quantization for approximate nearest neighbor search.
\newblock In {\em Proceedings ofthe 31th International Conference on Machine Learning (ICML}, volume~32, pages 838–846,, 2014.

\bibitem{NSDI2023}
Zili Zhang, Chao Jin, Linpeng Tang, Xuanzhe Liu, and Xin Jin.
\newblock Fast, approximate vector queries on very large unstructured datasets.
\newblock In {\em 20th USENIX Symposium on Networked Systems Design and Implementation (NSDI 23)}, pages 995--1011, Boston, MA, April 2023. USENIX Association.

\end{thebibliography}
\end{multicols}

\end{document}